\documentclass[aps,prx,twocolumn,amsmath,amssymb]{revtex4-2}
\usepackage{graphicx}
\usepackage[colorlinks=true,citecolor=blue,linkcolor=blue,urlcolor=blue]{hyperref}
\usepackage{amsmath,amssymb,bm,physics}
\usepackage[colorlinks=true,citecolor=blue,linkcolor=blue,urlcolor=blue]{hyperref}
\usepackage{comment}
\usepackage[caption=false]{subfig}
\newcommand{\lz}{L_z}                    % orbital Lz
\newcommand{\rr}{r^2}                    % radial square
\begin{document}

\title{Emergent nonlocal interactions induced by quantized gauge fields in topological systems}

\author{Adel Ali}
\affiliation{Department of Physics and Astronomy, Texas A\&M University, College Station, TX, 77843 USA}
\author{Alexey Belyanin}
\affiliation{Department of Physics and Astronomy, Texas A\&M University, College Station, TX, 77843 USA}

\date{\today}

\begin{abstract}

We study fermionic and bosonic systems coupled to a real or synthetic static gauge field that is quantized, so the field itself is a quantum degree of freedom and can exist in coherent superposition. A natural example is electrons on a quantum ring encircling a quantized magnetic flux (QMF) generated by a superconducting current. We show that coupling to a common QMF gives rise to an emergent interaction between particles with no classical analog, as it is topological and nonlocal (independent of interparticle distance). Moreover, the interaction persists even when the particles lie in a nominally field-free region, with the vector potential mediating the interaction. We analyze several one- and two-dimensional model systems, encompassing both real and synthetic gauge fields. These systems exhibit unusual behavior, including strong nonlinearities, non-integer Chern numbers, and quantum phase transitions. Furthermore, synthetic gauge fields offer high tunability and can reach field strengths that are difficult to realize with real magnetic fields, enabling engineered nonlinearities and interaction profiles.

\end{abstract}

\maketitle

\section{Introduction}

%name maybe: Quantum magnetostatics

The quantization of the electromagnetic (EM) field is a cornerstone of quantum optics and its quantum information applications, forming the theoretical bedrock of cavity quantum electrodynamics (QED) and, more recently, circuit QED, where quantized photon states coherently interact with matter \cite{Walther2006,BlaisRMP2021}.  

Here we explore the systems in which a \emph{static} magnetic field, real or synthetic (and its associated vector potential), is promoted to a quantum degree of freedom and can exist in a superposition of quantum states. This gives rise to a rich variety of interactions that have nonlocal and topological character.

Superconducting circuits provide one example of a natural arena where macroscopic persistent currents can occupy coherent superposition states.  The flux qubit experiments demonstrated superpositions of clockwise and counter-clockwise circulating currents, establishing that macroscopic magnetic flux states can also exhibit quantum coherence  \cite{Mooij1999,Friedman2000,Wal2000}. Since then, advances in materials, circuit design, and amplification techniques have pushed coherence times beyond the 0.1–1 ms range, while enabling high single-qubit gate fidelities  \cite{Kjaergaard2020}.  

In parallel, tremendous progress has been made in \emph{synthetic gauge fields} -- artificial EM potentials engineered for neutral particles \cite{review_gauge}. Pioneering experiments with ultracold atoms \cite{CooperRMP2019} and photons have emulated magnetic fields and spin-orbit couplings by tailoring laser phases and intensities. These advances allow neutral atoms or photons to exhibit Lorentz-force dynamics, Landau quantization, and topological band structures akin to electrons in real magnetic fields. 
Synthetic gauge fields are implemented in various ways across physical platforms. In cold atomic gases, laser-induced tunneling with spatially dependent phases can produce effective magnetic flux penetrating optical lattices. Using internal atomic states as extra lattice sites (a \emph{synthetic dimension}) has enabled realizations of the Harper–Hofstadter model on ribbon geometries, leading to direct observations of chiral edge currents in bulk-neutral atomic systems \cite{Mancini2015,Stuhl2015}. In photonics, dynamically modulated ring resonator lattices and waveguides have realized equivalent gauge potentials for light, giving rise to topologically protected photonic edge states and Landau levels in purely optical platforms \cite{Hafezi2013,Ozawa2019,photonics}. Superconducting-circuit networks of resonators and qubits, driven parametrically, have likewise been proposed and partially demonstrated as Hofstadter simulators with tunable magnetic flux per plaquette \cite{Roushan2017}.

While synthetic gauge fields are often implemented for classical fields (e.g., laser phases that are externally imposed), an exciting frontier is the use of \emph{quantized} gauge fields in these systems. In other words, instead of a fixed background gauge potential, one considers a dynamic global quantum field that can mediate interactions between particles. In traditional QED, it is well known that integrating out the quantized electromagnetic field yields effective interactions between charges or atoms in a cavity. In the context of artificial gauge fields, a QMF could similarly induce new types of photon-photon or atom-atom interactions that would not exist if the gauge field were classical.

Our goal is to merge these threads by quantizing the (real and synthetic) gauge fields and coupling it nonlocally to mobile charged or neutral particles.  This framework opens the door to studying hybrid light–matter many-body physics, flux-mediated interactions, and topological phase transitions in a regime that is both fundamentally important and experimentally accessible with current superconducting-circuit, cold-atom, or photonic technologies.

The structure of the paper is as follows. In Sec.~II we consider a one-dimensional (1D) particle system on a quantum ring (QR) and show that a single QMF threading a QR mediates an interaction between the particles that has topological and nonlocal properties. When the ring lies in a field-free region but nonzero vector potential from toroidal structure of a flux qubit or nearly field-free when the ring size is sufficiently larger than the planar flux qubit, the vector potential of the QMF will mediate the interaction demonstrating an example of a purely quantum effect where particles are coupled to each other with no real field associated with this interaction. We also analyze the systems with multiple QRs coupled to the same QMF. 

The nonlocality in these 1D systems is manifested as all-to-all long-range correlations of particle momenta, and the nontrivial topology is manifested as invariance under deformation of the ring and flux qubit.
The many-particle minimal-coupling Hamiltonian is exactly and analytically diagonalizable (no RWA), exhibiting a quantum phase transition from a zero-current ground state to a pair of degenerate chiral ground states.

We then turn to two-dimensional (2D) systems of particles that are known to exhibit topological character in the presence of a magnetic field. Here we investigate what happens when the magnetic flux is not a classical parameter in the Hamiltonian but a quantum degree of freedom (DOF). In Sec.~III we start with a one-particle problem and show that coupling to a quantum superposition of magnetic flux states results in the mixing of topologically distinct phases and the generalization of the Chern number to  noninteger  values. Continuing with a many-particle problem, we study the details of the QMF-mediated interaction between two particles  in Sec.~IV, and in a many-particle system in Secs.~V and VI. In particular, we show the possibility of quantum phase transitions in both bosonic and fermionic systems. In Sections VII and VIII we consider various models with artificial gauge fields, both in 1D and 2D. In particular, we consider a 2D lattice model which generalizes the Hofstadter model to a superposition of two flux states and shows a rich variety of metallic and insulating states. 

Section IX discusses possible implementations of artificial QMFs in the systems of superconducting qubits, waveguide and cavity QED, synthetic dimensions, and quantum lattices. The conclusions are in Sec.~X. 

Our findings bridge multiple areas of physics including condensed matter physics, quantum optics, and quantum information science, suggesting new routes to realize nonlocal topological interactions in a variety of experimental platforms, from superconducting circuit QED to cold atoms, ions, and photonic systems.

\begin{figure}[t]               % placement “t” for top; use h or b if preferred
  \centering
  %--- sub-figure (a) ---
  \subfloat[]{%
    \includegraphics[width=0.50\linewidth]{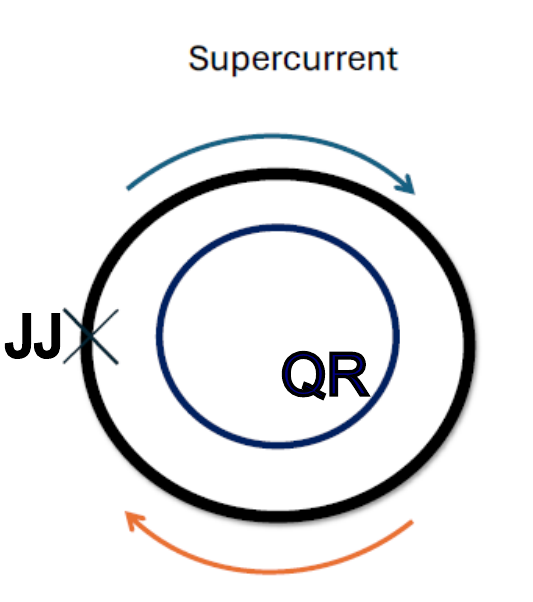}%
    %\label{fig:panel_a}
    }%
  \hfill
  %--- sub-figure (b) ---
  \subfloat[]{%
    \includegraphics[width=0.50\linewidth]{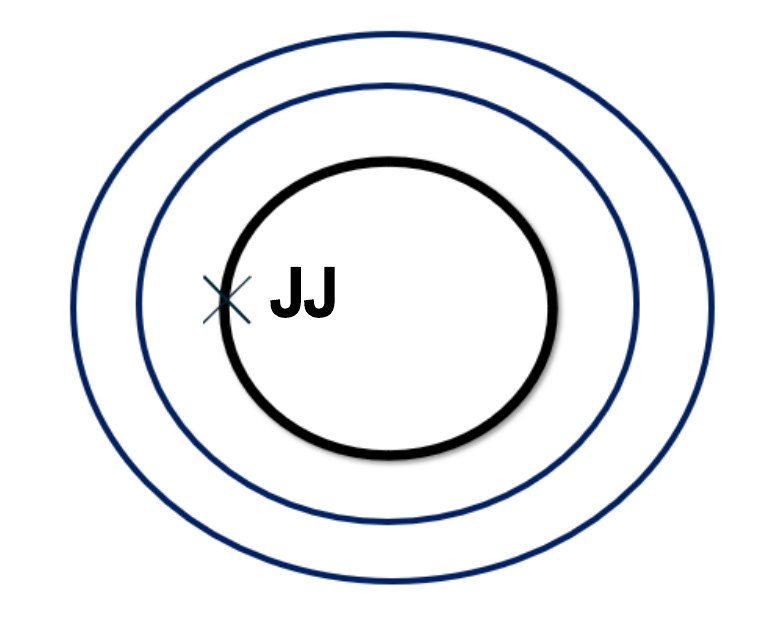}%
    %\label{fig:panel_b}
    }%
  \hfill
  %--- sub-figure (c) ---

  \caption{   
(a) Electrons on a QR (thin circle) coupled to the QMF created by the supercurrent loop (bold circle). JJ and a cross mark the Josephson junction;  (b)  Electrons on two separate QRs (thin circles) coupled nonlocally through the QMF created by the supercurrent loop (bold circle). QRs could be inside or outside the supercurrent loop, as long as they encircle a common magnetic flux.   }
    
  \label{fig1}
\end{figure}

\section{Particles on a quantum ring interacting via a quantized magnetic flux}

As a simple example highlighting the interaction between the particles mediated by the QMF, consider a quantum ring (QR) with charged particles, with a QMF threading the ring.

In the classical electrodynamics, EM interactions are mediated by ``real'' electric and magnetic fields produced by the charges. The Aharonov-Bohm (AB) effect shows that even in the absence of fields the classical EM vector potential can still affect the quantum state of a charged particle. Here we show that the quantized gauge potential gives rise to the interaction between the particles, which is topological and nonlocal. It should be noted that this type of coupling mediated by a QMF will exist whether or not the particles are in the field-free region.

As sketched in Fig.~\ref{fig1}(a),  consider two particles of mass $m$ confined to a one-dimensional ring of radius $R$, which is located inside a a loop of supercurrent supporting a QMF. We will also assume that there is a Josephson junction in the supercurrent loop which makes the energy levels of the QMF non-equidistant, and we will take into account only the lowest two levels separated by energy gap $\Delta$. This is a situation typically considered for flux qubits of any architecture.   This is not essential for any of the effects described in the paper, but having the QMF as a two-level system makes derivations easier and the results more transparent.

For two nonrelativistic particles constrained to move on a 1D ring of radius $R$ and in the absence of any gauge field, each particle has momentum $p_{i}$ (with $i=1,2$) quantized in units of $2\pi \hbar / R$ due to periodic boundary conditions. When we introduce a QMF inside the ring, the minimal coupling for each charged particle is achieved via the substitution $p_i \to p_i - q A$ in the kinetic energy, where $q$ is an effective charge and $A \to \hat{n}A$ is the vector potential of the QMF which is now an operator: $\hat{n} = (1+\sigma^z)/2$ projects onto the flux-occupied state, where 
$\sigma^x$ and $\sigma^z$ are Pauli matrices acting on the flux qubit space.

\begin{figure}[t]               % placement “t” for top; use h or b if preferred
  \centering
  %--- sub-figure (a) ---
  \subfloat[]{%
    \includegraphics[width=0.50\linewidth]{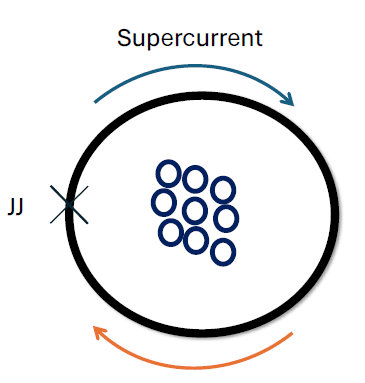}%
    %\label{fig:panel_a}
    }%
  \hfill
  %--- sub-figure (b) ---
  \subfloat[]{%
    \includegraphics[width=0.50\linewidth]{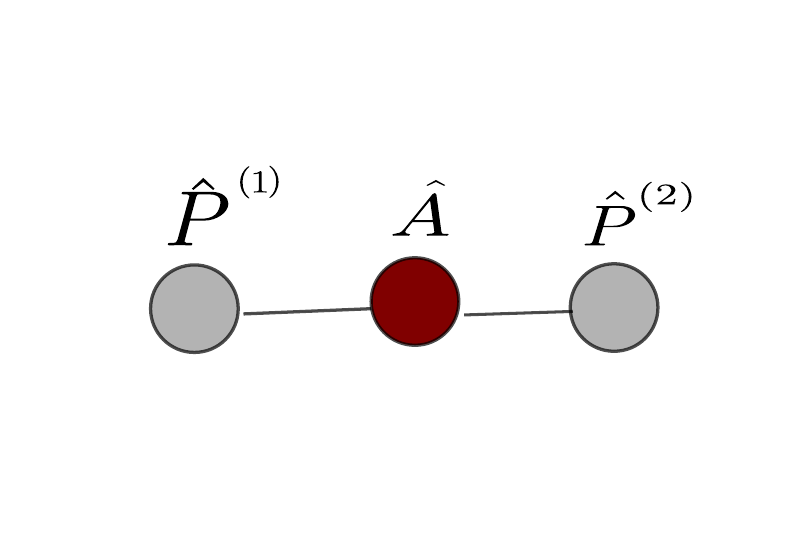}%
   % \label{fig:panel_b}
    }%
  \hfill
  %--- sub-figure (c) ---

  \caption{    
   (a) Electrons on different QRs coupled to the QMF created by the flux qubit. 
 (b) Schematic of the interaction between particle momenta mediated through the vector potential.}
    
  \label{fig2}
\end{figure}

The Hamiltonian of the coupled system can be written as:
\begin{equation}
H = \sum_{i=1}^2 \frac{1}{2m}\big(\hat{p}_i - q\hat{A}\big)^2 + \Delta \sigma_x \,,
\label{eq:H_ring}
\end{equation}
where $\Delta \sigma_x$ is the Hamiltonian for the flux qubit in the basis of clockwise and counter clockwise current.
To analyze the interaction between the particles and QMF, one expands the square to find $\sum_i \frac{\hat{p}_i^2}{2m} + \frac{q^2 \hat{A}^2}{2m} - \frac{q}{m} \hat{p}_i \hat{A} $. The term $-\frac{q}{m}(\hat{p}_1 + \hat{p}_2)\hat{A}$ describes the interaction between the particles' motional DOF and the QMF, sketched in Fig.~\ref{fig2}(b), while the $A^2$ term (often called the diamagnetic term) represents the energy cost of the gauge field when excited. Importantly, if one of the particles gains momentum $p$, the $pA$ coupling can transfer quanta of excitation between the particles and the gauge field.

Before tackling the exact Hamiltonian, one can gain insight and intuitive physical picture into the nature of the coupling by studying the dispersive regime which is also realistic for experiments. In the dispersive regime where the QMF mode is far detuned, i.e. \( \Delta \gg g\), where $g$ is defined below, the QMF remains mostly in its ground state and one can adiabatically eliminate the QMF DOF. In the second-order perturbation theory, integrating out $\hat{A}$ yields an effective interaction between the two particles. The virtual transitions between QMF levels impart the coupling between momenta of the particles, correlating their motion. The effective Hamiltonian for the particles alone becomes, up to the first order in \( 1/\Delta \), 

%{\bf if this is for two rings, there will be different $g_1$ and $g_2$, still one ring}
\begin{align}
\hat{H}_{\text{eff}} =\; &
\left( \frac{g}{\hbar^2} - \frac{g^2 \phi^2}{2\Delta \hbar^2} \right)(\hat{p}_1^2 + \hat{p}_2^2) \nonumber \\
& + \left( -\frac{g\phi}{\hbar} + \frac{g^2 \phi^3}{2\Delta \hbar} \right)(\hat{p}_1 + \hat{p}_2) \nonumber \\
& - \frac{g^2 \phi^2}{\Delta \hbar^2} \hat{p}_1 \hat{p}_2
+ \left( \frac{g\phi^2}{2} - \frac{g^2 \phi^4}{8\Delta} \right). 
\label{eq:Heff}
\end{align}
Here $\phi = q\Phi/(2\pi\hbar)$ is the dimensionless magnetic flux and we introduced the characteristic energy scale $g = \frac{\hbar^2}{2mR^2}$,
corresponding to the kinetic energy of a particle on the ring. Its value is determined by the ring material and size. 

The first term is a renormalized kinetic energy for each particle due to the QMF coupling, which effectively increases the electron mass. The second term describes the flux screening. The last term is a constant energy shift that can be dropped. The most interesting term is the nonlocal interaction \( \hat{p}_1 \hat{p}_2 \)  which tends to align the particles momenta. This is different from the typical EM interaction which depends on the  distance between the interacting particles. Moreover, for a toroidal QMF \cite{PhysRevA.110.022604} the particles on the ring can be located entirely in a field-free region. Note that in a standard cavity QED, the situation with multiple atoms coupled to a single mode in a cavity would not meet the nonlocality requirement in the same way since, hypothetically moving the atoms in the cavity, their coupling strength to a cavity field will change affecting the effective interaction strength between them.  

Two charged particles will have also an orbit-orbit interaction of the form \[
\hat V_{\mathrm{}}
= -\,\frac{q^{2}}{16\pi\varepsilon_{0} c^{2} R^{3}}\,
\frac{1+3\cos(\theta_{1}-\theta_{2})}{
\left|\sin\!\left(\tfrac{\theta_{1}-\theta_{2}}{2}\right)\right| }\;
\frac{1}{m^{2}}\!\left(\hat L_{1}\hat L_{2}\right),
\]
where the angles $\theta_{1,2}$ determine the positions of the particles. However this interaction is much weaker and is geometry-dependent.
%Here the coupling depends only on the enclosed flux by the ring     

For many particles on a QR there will be an all-to-all attraction leading to self-focusing in momentum space. 
This shows how virtual excitations of a QMF can generate long-range interactions among otherwise free particles, and may induce flux screening and momentum-polarized phases for larger \( N \). Higher orders of the expansion will have more terms that couple the momenta and renormalize the higher powers. 

This interaction has no classical analogue as opposed to the AB effect. If there are two different concentric rings as in Fig.~\ref{fig1} (b),  with one electron on one of them, the introduction of another electron in the second ring (for example, by means of gating) will face a nonlocal blockade that depends on the state of the first electron. 
This effect persists even when $m=0$ for both particles, so there is no inductive coupling, and for arbitrary different radii of the two rings.

The full Hamiltonian can be exactly diagonalized to find polariton-like eigenstates and look at hints for phase transitions. For $N$ particles it reads
\begin{equation}
\hat{H} = \sum_{i=1}^{N} g\left( -i\frac{\partial}{\partial \theta_i} - \hat{n} \phi \right)^2 + \Delta \sigma^x,
\label{eq:Hfull}
\end{equation}
where $\theta_i$ is the angular coordinate of the $i$-th particle.

In the basis of angular momentum states 
$$\left\langle\varphi \mid m\right\rangle=\frac{e^{-i m \theta}}{\sqrt{2 \pi R}}$$
labeled by quantum number $m$ (not to be confused with particle mass which is absorbed inside $g$), the Hamiltonian becomes 
\begin{equation}
H=\sum_{i=1}^{N}g\left(m_{i}- \hat{n}\phi^{}\right)^{2}+ \Delta \sigma_{x}^{}. 
\nonumber 
\end{equation}

In the second quantization picture,  
\begin{equation}
H=\sum_{m}^{}\left(m_{i}- \hat{n}\phi^{}\right)^{2}\hat{c}_m^\dagger \hat{c}_m + \Delta \sigma_{x}^{}, 
\label{eq2}
\end{equation}
where $\hat{c}_m^\dagger$ and $\hat{c}_m$ are either bosonic or fermionic creation and annihilation operators of particle $m$-states. Both cases will be considered later. 

%we set $g \equiv1$ so $\Delta$ is now in units of g. typically $\Delta$ is few to tens of GHZ while g for micron radius is around 10 MHZ

Of course this model Hamiltonian neglects Coulomb interaction between electrons which is always there and has to be added for any realistic modeling. However, the coupling we want to highlight here is independent of the Coulomb effects and moreover it is of topological nature, meaning that it does not have the geometric dependence of Coulomb's law, $\frac{1}{|r_i-r_j|}$. 

The Hamiltonian in Eq.~(\ref{eq2}) has a characteristic nonlinearity (the product of more than two operators) originated from the first term.  This nonlinearity can be used, e.g., to make a blockade, where the existence of an electron can block other electrons from entering the same ring or, in the case of different concentric rings, one electron can block the electrons in other rings in a nonlocal way. The nonlocal blockade can also happen, for example, in a system of two quantum rings coupled through a toroidal flux qubit \cite{PhysRevA.110.022604}. 

\subsection{Exact diagonalization at fixed particle configuration}

This model can be diagonalized analytically since the wavefunctions of particles $e^{-im\theta}$ on the ring are independent of the flux state, unlike the 2D case in the subsequent sections.

Since $\hat n= \displaystyle \tfrac{1-\sigma^z}{2}$ and with $\hat n_m\equiv \hat c_m^\dagger \hat c_m$, we write
\begin{align}
H&=\sum_m\!g\left[\Big(m-\tfrac{\phi}{2}\Big)^2+\tfrac{\phi^2}{4}\right]\hat n_m
+g\phi\,\sigma^z\!\sum_m\!\Big(m-\tfrac{\phi}{2}\Big)\hat n_m \notag\\
&\quad+\;\Delta\,\sigma^x
\;\equiv\; H_0+\phi\,K\,\sigma^z+\Delta\,\sigma^x .
\label{5}
\end{align}
Here 
\begin{equation}
H_0=\sum_m \Big(m-\tfrac{\phi}{2}\Big)^2 \hat n_m + \tfrac{\phi^2}{4}\,\hat N,
\qquad
K=\sum_m \Big(m-\tfrac{\phi}{2}\Big)\hat n_m,
\nonumber 
\end{equation}
and $\hat N=\sum_m \hat n_m$. Since $[H_0,K]=0$ and $K$ acts only on the particles, $K$ is a good quantum number. The coupling here is identical for all particles if they are on the same ring as it only depends on $\phi$, in contrast to a system of emitters in cavity QED where the coupling depends on their position with respect to a cavity field distribution. 

\subsubsection{Block reduction and QMF energy}

In any joint eigenstate $\ket{\{n_m\}}$ of all $\hat n_m$ with
$K\ket{\{n_m\}}=k\ket{\{n_m\}}$, the Hamiltonian reduces to a
$2\times 2$ problem
\begin{align}
H(\{n_m\}) &= E_0(\{n_m\}) 
  + \big(\Delta\,\sigma^x + \phi k\,\sigma^z\big), \nonumber 
  \\[4pt]
E_0(\{n_m\}) &= \sum_m \Big(m-\tfrac{\phi}{2}\Big)^2 n_m
  + \tfrac{\phi^2}{4}\,N . \nonumber 
\end{align}
Notice that the $\phi k$ term acts as a pseudo-$z$-direction magnetic field for the qubit which can be used to control the qubit state. The eigenvalues are $\pm r_k$ with
\begin{equation}
r_k=\sqrt{\Delta^2+\phi^2 k^2}. \nonumber 
\end{equation}
Thus the exact energy in that particle sector is
\begin{equation}
E(\{n_m\})=E_0(\{n_m\})-r_k . \nonumber
\end{equation}

\subsection{Exact ground-state spinor}

In the $\sigma^z$ basis $\{\ket{\uparrow},\ket{\downarrow}\}$ the qubit
ground spinor (eigenvalue $-r_k$) is
\begin{align}
\ket{q(k)} &=
\frac{\Delta\,\ket{\uparrow}-\big(r_k+\phi k\big)\ket{\downarrow}}
{\sqrt{\Delta^2+\big(r_k+\phi k\big)^2}} \nonumber \\
&= \cos\!\frac{\theta_k}{2}\,\ket{\downarrow}
   - \sin\!\frac{\theta_k}{2}\,\ket{\uparrow},
\end{align}
with
\begin{equation}
\cos\theta_k=\frac{\phi k}{r_k}, \; 
\sin\theta_k=\frac{\Delta}{r_k}. \nonumber 
\end{equation}
Therefore, the exact ground state of the full system is a product in a
single $k$-sector,
\begin{equation}
\ket{\Psi_{\mathrm{GS}}}=\ket{\{n_m\}^\ast}\otimes \ket{q(k^\ast)},
\end{equation}
where $\ket{\{n_m\}^\ast}$ minimizes $E(\{n_m\})$.

%%%%%%%%%%%%%%%%%%%%%%%%%%%%%%%%%

\subsection{Phase transitions}

\subsubsection{Bosons at $T=0$}

For bosons the minimum energy at fixed integer $m$ is reached when all $N$ particles are in
one mode, $n_m=N$, $n_{m'\neq m}=0$, hence
$k=N(m-\tfrac{\phi}{2})$. The lower branch of energy is
\begin{align}
E_B(m) &= N\Big(m-\tfrac{\phi}{2}\Big)^2+\tfrac{\phi^2}{4}N \nonumber \\
&\quad - \sqrt{\Delta^2+\phi^2 N^2\Big(m-\tfrac{\phi}{2}\Big)^2}.
\nonumber 
\end{align}
If we pick $m_0=\arg\min_{m\in\mathbb{Z}} E_B(m)$ and
$k_0=N(m_0-\tfrac{\phi}{2})$, then the ground state becomes 
\begin{equation}
\ket{\Psi_{\mathrm{GS}}}=
\frac{(\hat c_{m_0}^\dagger)^N}{\sqrt{N!}}\ket{0}
\;\otimes\;\ket{q(k_0)}. 
\nonumber 
\end{equation}
A finite-size bifurcation to the ground states with two current-carrying minima
occurs at
\begin{equation}
\Delta_c^{}=\frac{g\phi^2}{2}\,N ,
\label{deltac}
\end{equation}
yielding two symmetry-related ground states with $\pm k_0\neq 0$ for
$\Delta<\Delta_c^{(B)}$.

%Notice that in the thermodynamic limit the coupling is normalized by $\sqrt{N}$, plugging g back will give a critical coupling $\phi=\sqrt{g\Delta}$

There is a global parity symmetry in the Hamiltonian of Eq.~\eqref{5} with respect to changing the sign of K and $\sigma_z$. For the particles subsystem, Eq.~\eqref{deltac} is the critical value for a crossover (bifurcation). For a phase transition going to the thermodynamic limit, one has to renormalize the coupling $\phi$ and the order parameter by $\sqrt{N}$, so that $\Delta_c^{}=\frac{g\phi^2}{2}$, or allow extensive $\Delta$ (using an ensemble of qubits $N\Delta$). 

This phase transition is structurally similar to the Dicke superradiant phase transition, but here the diamagnetic $A^2$ term of the minimal coupling is not ignored. 

One possible implementation of such bosonic systems with real magnetic fields could be multiple QRs in Fig.~\ref{fig2}(a) which form the hard-core bosons. 
%When $N$ is too large, the two level approximation of the flux qubit breaks down and the leakage to higher states of the qubit will happen. 

\subsubsection{``Spinless'' fermions at $T=0$}

The crossover exists also for a system of fermions, where we ignored the spin term in the Hamiltonian for simplicity.  
For a fixed number \(N\) of fermions and fixed \(k\), the fermionic configuration minimizing \(E_0\) is a contiguous block
\(
\mathcal{S}_M=\{M,M{+}1,\dots,M{+}N{-}1\}
\)
centered as close as possible to \(x\equiv\phi/2\).
Define
\(
A=\sum_{i=0}^{N-1}i^2=\frac{(N-1)N(2N-1)}{6}
\),
\(
B=\sum_{i=0}^{N-1}i=\frac{N(N-1)}{2}
\).
For a given configuration \(\mathcal{S}_M\),
\begin{align}
k(M)&=\sum_{m\in\mathcal{S}_M}(m-x)=N(M-x)+B, \nonumber \\
E_0(\mathcal{S}_M)
&=
%\sum_{m\in\mathcal{S}_M}(m-x)^2+\tfrac{\phi^2}{4}N
\Big[A-\frac{B^2}{N}\Big]+\frac{k(M)^2}{N}+\tfrac{\phi^2}{4}N. \nonumber
\end{align}
Therefore the problem reduces to minimizing over the \emph{single} integer \(M\) (equivalently over the discrete set of \(k\)-values)
\begin{equation}
{\;\mathcal{E}(k)=C_N+\frac{k^2}{N}-\sqrt{\Delta^2+\phi^2 k^2}\;},\; 
C_N=A-\frac{B^2}{N}+\tfrac{\phi^2}{4}N, 
\label{eq:Ek}
%\nonumber 
\end{equation}
with \(k\equiv k(M)=N(M-\tfrac{\phi}{2})+B\). \\

{\it The critical point in the continuum limit.}
Treating \(k\) as continuous gives
\(
\partial_k\mathcal{E}=0
\Rightarrow
\frac{2k}{N}-\frac{\phi^2 k}{\sqrt{\Delta^2+\phi^2 k^2}}=0
\).
Besides \(k=0\), nonzero solutions satisfy
\(
\sqrt{\Delta^2+\phi^2 k^2}=N\phi^2/2
\),
hence
\begin{equation}
k_\ast^2=\frac{N^2\phi^2}{4}-\frac{\Delta^2}{\phi^2},
\; 
{\;\phi_c^2=\frac{2\Delta}{N}\;}. \nonumber 
\end{equation}
Near \(\phi_c\), we can approximate 
\(
-\sqrt{\Delta^2+\phi^2 k^2}\simeq -\Delta-\frac{\phi^2}{2\Delta}k^2
\),
so the quadratic coefficient
\(
\frac{1}{N}-\frac{\phi^2}{2\Delta}
\)
changes sign at \(\phi_c\).

For \(\phi\) fixed, one should select the integer \(M^\star\) (hence \(k^\star=k(M^\star)\)) minimizing energy in Eq.~\eqref{eq:Ek}.

For even \(N\), a block symmetric about \(x=\phi/2\) gives \(k=0\) for \(|\phi|<\phi_c\). For \(|\phi|>\phi_c\) two symmetry-related minima \(\pm k^\star\) appear (Ising-like \(k\!\to\!-k\)).
For odd \(N\), the optimal block must include \(\operatorname{round}(x)\), leaving a small residual
\(k_0=\operatorname{round}(x)-x\in[-\tfrac12,\tfrac12]\);
for \(|\phi|>\phi_c\) the pair \(\pm k^\star\) again minimizes \(\mathcal{E}\).

Let \(\mathcal{S}_{M^\star}\) be the minimizing block and \(k^\star=k(M^\star)\).
Then the total ground state factorizes as
\begin{equation}
\ket{\mathrm{GS}}=\ket{\mathcal{S}_{M^\star}}\otimes\ket{q(k^\star)},
\; {\rm } \; 
\ket{q(k)}=\cos\!\frac{\vartheta_k}{2}\ket{\!\downarrow_z}-\sin\!\frac{\vartheta_k}{2}\ket{\!\uparrow_z}.  \nonumber 
\end{equation}

The order parameter \(k\) is zero for \(|\phi|<\phi_c\) (even \(N\)), and takes \(\pm k^\star\) for \(|\phi|>\phi_c\).

So far we assumed a uniform $\phi$ or vector potential $\hat{A}$ which will induce forward scattering only, described by the interaction term $V_{\rm eff}\propto -K^2$ where $K$ is defined in Eq.~\eqref{5}. If $\hat{A}(\theta)$ has a structure (or some harmonic mode) then momentum transfer will be allowed and one can get channels for attractions akin to Cooper pairing. In particular, $A(\theta)=\sum_q A_q e^{iq\theta}$ yields $V_{\rm eff}\propto -\sum_{q,q'} \phi_q \phi_{q'}, j_q j_{q'}$ with finite-$q$ pair scattering that enables pair density wave (Amperean) superconductivity. 
If the source of the vector potential is another SC loop, to have a uniform $A$ an axial symmetry is needed. if the source is a trapped flux, i.e., the field-free region, the vector potential is always uniform for any geometry.

%%%%%%%%%%%%%%%%%%%%%%%%%%%%%%%%%%%%%%%%%%%%%%%%%%%

\section{2D electron gas coupled to a QMF}

\subsection{One-particle Hamiltonian}

\begin{figure}[t]
\includegraphics[width=5cm]{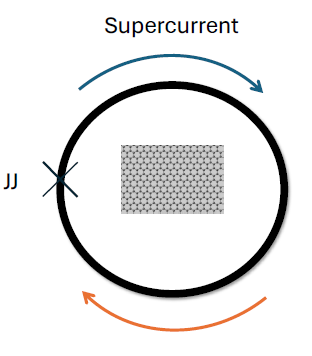}
\centering
\caption{ A schematic of the 2DEG in the QMF of a flux qubit.}
\label{figure3}
\end{figure}

Now let us consider a more complicated system: a 2D gas of charged particles coupled to a QMF, as sketched in Fig.~\ref{figure3}. We will call it a 2D electron gas (2DEG) for simplicity, but the same treatment could apply to any bosonic or fermionic system coupled to a real or artificial QMF. 

For the Landau problem of non-interacting particles with charge $q$ confined 
%to an area $L_xL_y$ 
in the x-y plane and with uniform magnetic field along the z-axis, defined in the Landau gauge by the vector potential  $\mathbf {A} = (0,Bx,0)$, the single-particle Hamiltonian reads
\begin{equation}
 \hat{H}={\frac {1}{2m}}\left({\hat {\mathbf {p} }}-q{\hat {\mathbf {A} }}\right)^{2}={\frac {{\hat {p}}_{x}^{2}}{2m}}+{\frac {1}{2m}}\left({\hat {p}}_{y}-qB{\hat {x}}\right)^{2}. 
 \label{27}
 \end{equation}
Here we assumed as before that the magnetic flux is quantized, so that we can promote $B$ into $\hat{n}B$ with $\hat{n}$ being a quantum operator acting on the two states of the magnetic flux $\hat{n}|0\rangle=0$, $\hat{n}|1\rangle=|1\rangle$, and with  energy separation $\Delta$ between these two states.  The single-particle Hamiltonian could be then written as 
\begin{equation}
\hat {H}={\frac {{\hat {p}}_{x}^{2}}{2m}}+{\frac {1}{2m}}\left({\hat {p}}_{y}-q\hat{n}B{\hat {x}}\right)^{2}+\Delta\sigma^{x}. 
\label{28}
\end{equation}
For the first two terms in the Hamiltonian combined, which we will denote as $H_1$, $$|e^{i(k_{y}y)}\phi _{n}(x-x_{0})\rangle\otimes|1\rangle $$ is an eigenstate with energy $E_{n}=\hbar \omega _{\rm {c}}\left(n+{\frac {1}{2}}\right),\quad n\geq 0$ where $|\phi _{n}(x-x_{0})\rangle$ are the harmonic oscillator eigenstates, $\omega_c = eB/m$ is the cyclotron frequency, whereas  $$|e^{i(k_{y}y+k_{x}x)}\rangle\otimes|0\rangle$$ is an eigenstate with energy $E_{k}=\frac{\hbar^2 k^{2}}{2m}$. For a flux qubit the magnetic field strength can reach mT around the center of the qubit. 
%\textcolor{red}{\bf define $\phi_n$}
%for a 2DEG sample placed in the center optical probe is possible 

Since the QMF can be in either a zero or nonzero flux state (or their superposition), the eigenenergy  bands will have an alternating Chern number between 0 and 1, which creates a peculiar system of two topologically distinct bands. For example, an electron  will have a protected chiral edge state or will backscatter, depending on its wavevector.

\begin{figure}[t]               % placement “t” for top; use h or b if preferred
  \centering
  %--- sub-figure (a) ---
  \subfloat[]{%
    \includegraphics[width=0.50\linewidth]{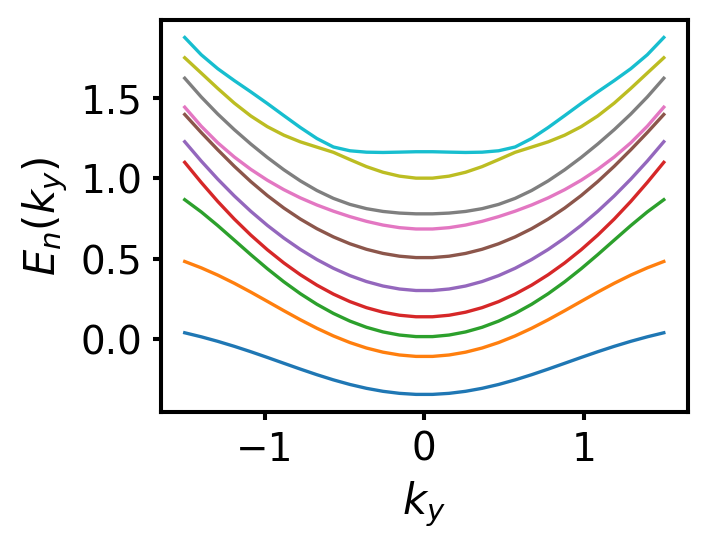}%
    \label{fig:panel_a}}%
  \hfill
  %--- sub-figure (b) ---
  \subfloat[]{%
    \includegraphics[width=0.450\linewidth]{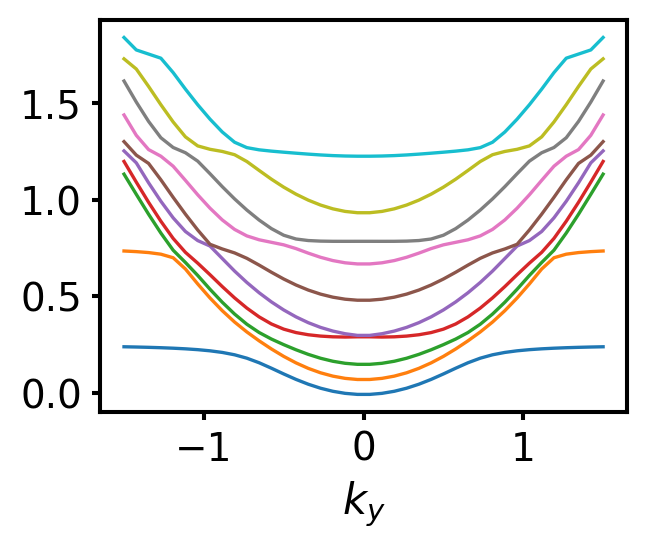}%
    \label{fig:panel_b}}%
  \hfill
  %--- sub-figure (c) ---
 %\subfloat[]{%
    %\includegraphics[width=0.50\linewidth]{.png}%
    %\label{fig:panel_b}}%
  %\hfill

  \caption{    
    Exact diagonalization of the one particle problem in the Landau gauge. All energies $E_n$ and $\Delta$ are normalized by $\hbar \omega_c$. (a) $\Delta =1$,  
    (b) $\Delta=1/5$.}
    
  \label{fig4:combined}
\end{figure}

The term $\Delta\sigma^{x}$ makes the system not instantly diagonalizable. 
The numerical solution of its eigenvalue problem is illustrated in Fig.~\ref{fig4:combined} for different values of $\Delta$. In this two-level approximation the state of the field can be represented as a pseudospin 1/2 field. For small $\Delta$, the degeneracy of $H_1$ at the points $ k_y =\sqrt{\frac{m}{\hbar}(\omega _{\rm {c}}\left(2n+1\right)+4\Delta)}$, will be lifted. In 2D momentum space the paraboloid of free-electron energy is intersected by equally spaced lines of Landau levels. 

The many-body picture will be our primary interest in the next sections. It is not straightforward since there is a collective coupling between the particles and the QMF and accordingly the nonlinearity. This will result in the coupling and anticrossing of the two topologically distinct states around the degeneracy points. 
As $\Delta$ gets larger, the coupling between the two types of states will increase. 

The energy $\Delta$ can be tuned by the parameters of the circuit including inductance and Josephson junction charging energy; it can vary from $\sim 1$ MHz for fluxonium qubits \cite{fluxoiummhz} to GHz in flux qubits. The practical challenges will come from the weak strength of the magnetic field corresponding to the QMF: in the  mT range for conventional flux qubits. This strength depends on the size of the qubit and superconducting critical current. Observing Landau quantization at such field requires 
extremely clean 2D materials and mK temperatures.  
However, the same framework and concepts can be applied for synthetic gauge fields that are tunable and can be implemented in various material systems.

In the limit of strong coupling (high $\omega_c$), the quantized flux states and particles become  entangled, and one enters a regime similar to cavity QED and circuit QED in the ultrastrong and deep coupling limit, because the coupling energy is not small as compared to eigenenergies of uncoupled subsystems. The resulting bound states mediated by exchange of discrete flux quanta pave the way to the same kind of exciting physics envisaged for photonic systems and their superconductivity analogs in circuit QED systems.

Since the magnetic flux is a dynamical degree of freedom, a topological transition between the two topologically distinct phases can be coherently controlled, e.g., with microwave photons.

\begin{figure}[t]
\includegraphics[width=6cm]{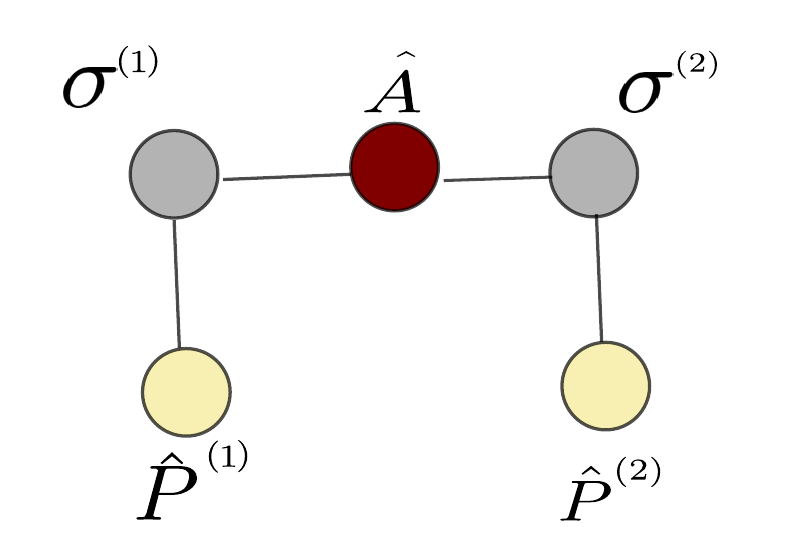}
\centering
\caption{ A sketch of the interaction between momenta of massless particles in graphene mediated through the sublattice pseudospin.}
\label{pseudospin}
\end{figure}

A similar system that could also be described by the single-particle  Hamiltonian similar to  Eq.~\eqref{28} is a bilayer system where for each layer there is different magnetic field and there is a tunneling $\Delta$ between the two layers. Perhaps the most straightforward example is bilayer graphene with the two layers experiencing different strains. Of course electrons in graphene are relativistic and the Hamiltonian needs to be modified appropriately. 
%{\bf can you add here the case of graphene in more detail? Otherwise it is better ro remove this paragraph. }
For example, the Hamiltonian for two non-interacting electrons in graphene under a perpendicular magnetic field \( \mathbf{B} = B\hat{z} \), using minimal coupling \( \mathbf{p}_i \to \mathbf{p}_i + e\mathbf{A}(\mathbf{r}_i) \), is given by 
\begin{equation}
\hat{H} = \hbar v_F \left[ \boldsymbol{\sigma}_1 \cdot \left( \mathbf{p}_1 + e\mathbf{A}(\mathbf{r}_1) \right) + 
\boldsymbol{\sigma}_2 \cdot \left( \mathbf{p}_2 + e\mathbf{A}(\mathbf{r}_2) \right) \right]
\nonumber 
\end{equation}
where \( \boldsymbol{\sigma}_i = (\sigma_x^{(i)}, \sigma_y^{(i)}) \) are Pauli matrices acting on sublattice pseudospin of particle \( i \), \( \mathbf{A}(\mathbf{r}) \) is the vector potential corresponding to the magnetic field \( \mathbf{B} \), \( v_F \) is the Fermi velocity in graphene. For a QMF, \( \mathbf{A}(\mathbf{r}) \)  becomes \( \hat{n} \mathbf{A}(\mathbf{r}) \).  
This gives rise to the coupling between the particles pseudospin and the coordinates, and indirectly, the momenta.
For N particles, in the symmetric gauge \( \mathbf{A} = \frac{B}{2}(-y, x) \), the Hamiltonian becomes:
\begin{equation}
\hat{H} = \hbar v_F \sum_{i=1}^{N} \boldsymbol{\sigma}_i \cdot \left( \mathbf{p}_i + \frac{eB\hat{n}}{2}(-y_i, x_i) \right)
\end{equation}
We get a collective momentum interaction for the massless particles mediated by the vector potential through the pseudospin, as sketched in Fig.~\ref{pseudospin}.  However, the focus in the rest of the paper will be on massive particles. 

\subsection{Non-integer Chern number}

Since a general state of the system is a superposition of states with zero and nonzero magnetic fluxes, the resulting mixing of two distinct topological phases gives rise to an arbitrary Chern number between zero and one. For example, for an entangled state of the type
$$\alpha|e^{i(k_{y}y+k_{x}x)}\rangle\otimes|0\rangle+\beta|e^{i(k_{y}y)}\phi _{n}(x-x_{0})\rangle\otimes|1\rangle $$

the Chern number would be $\beta^2$ since the first part of the wavefunction has trivial topology so it gives zero contribution.  

Another way to realize similar physics would be to have two tunneling-coupled systems with different real or artificial quantized gauge fields. One example would be two stacked 2DEG systems or two tunneling-coupled QRs, each  coupled to a different real magnetic flux or an artificial/synthetic gauge field.

In the discussion above we focused on the case where the magnetic field is described by the operator $\hat{n} B$.  It describes the hybridization of free and Landau-quantized electrons, or two topologically distinct states \cite{hybridtopology}.  However, there is also the possibility of having the magnetic field as a superposition of the kind $B_{0}+\hat{n} B$, i.e., biased by a classical field. Such a system  
will have different properties because it describes the hybridization of two Landau-quantized systems. Time reversal symmetry is completely broken in this case.

%%%%%%%%%%%%%%%%%%%%%%%%%%%%%%%%%%%%%%%%%%%%%%%%%%%%%

\section{Two particles coupled to a quantized magnetic flux}

So far we dealt with the 2DEG in a single-particle picture. To get a sense of many-body effects induced by coupling to a QMF, in this section we 
consider a system of two charged particles moving in two dimensions, each coupled to a shared quantized magnetic flux modeled as a two-level system (pseudospin). The total Hamiltonian reads 
\begin{equation}
\hat{H} = \sum_{i=1}^2 \left[ \frac{\hat{p}_x^{(i)2}}{2m} + \frac{1}{2m} \left( \hat{p}_y^{(i)} - q \hat{n} B \hat{x}^{(i)} \right)^2 \right] + \Delta \sigma^x,
\end{equation}
where \( \hat{n} \in \{0, 1\} \) is the occupation operator of the flux mode, and \( \Delta \sigma^x \) provides coherent tunneling between the two flux states.

%%%%%%%%%%%%%%%%%%%%%

We aim to perform a Schrieffer-Wolff transformation. 
Using the identity \( \hat{n} = \frac{1}{2}(I + \sigma^z) \), we define the operator
\begin{equation}
F = -\frac{qB}{m}(\hat{x}^{(1)} \hat{p}_y^{(1)} + \hat{x}^{(2)} \hat{p}_y^{(2)}) + \frac{q^2 B^2}{2m}(\hat{x}^{(1)2} + \hat{x}^{(2)2})
\end{equation}
which allows the Hamiltonian to be written as
\begin{equation}
\hat{H} = H_{\mathrm{kin}} + \frac{1}{2}F + \frac{1}{2}\sigma^z F + \Delta \sigma^x,
\nonumber 
\end{equation}
where \( H_{\mathrm{kin}} = \sum_{i=1}^2 \frac{\hat{p}_x^{(i)2} + \hat{p}_y^{(i)2}}{2m} \) is the kinetic energy.

%%%%%%%%%%%%%%%%%%%%%%%%%%%

\subsection{Adiabatic elimination of the flux pseudospin}

Assuming the dispersive limit with $\Delta$ much greater than the kinetic energy, similarly to the  QED seminal work \cite{dispersive},  we eliminate the pseudospin degree of freedom perturbatively. The ground state \( |g\rangle \) of \( \sigma^x \) satisfies \( \langle g|\hat{n}|g\rangle = 1/2 \).

\paragraph{First-order contribution.} The leading-order effective Hamiltonian becomes
\begin{equation}
H_{\mathrm{eff}}^{(1)} = H_{\mathrm{kin}} + \frac{1}{2}F,
\nonumber 
\end{equation}
which corresponds to each particle experiencing an average magnetic field of strength \( B/2 \):
\begin{equation}
H_{\mathrm{eff}}^{(1)} = \sum_{i=1}^2 \left[ \frac{p_x^{(i)2} + p_y^{(i)2}}{2m} - \frac{qB}{2m} x^{(i)} p_y^{(i)} + \frac{q^2 B^2}{4m} x^{(i)2} \right].
\nonumber 
\end{equation}

\paragraph{Second-order correction.} Virtual excitations of the pseudospin to its excited state induce an effective interaction. The second-order contribution is given by
\begin{equation}
H_{\mathrm{eff}}^{(2)} = -\frac{1}{8\Delta} F^2.
\nonumber 
\end{equation}

Expanding \( F^2 \) yields two-particle interaction terms, notably, 
\begin{align}
H_{\mathrm{int}}^{(2)} &= -\frac{q^2 B^2}{4m^2 \Delta} x^{(1)} p_y^{(1)} x^{(2)} p_y^{(2)} - \frac{q^4 B^4}{16 m^2 \Delta} \left(x^{(1)} x^{(2)}\right)^2 \nonumber \\
&  +\frac{q^3B^3}{m^2\Delta} \left[x^{(1)} \left(x^{(2)}\right)^2\, p_{y}^{(1)} + x^{(2)} \left(x^{(1)} \right)^2\, p_y^{(2)}\right] 
\nonumber 
\end{align}
+ single-particle terms.

%%%%%%%%%%%%%%%%%%%%%%%

\subsection{The effective Hamiltonian}

Combining both orders, the effective Hamiltonian reads:
\begin{align}
H_{\mathrm{eff}} &\approx \sum_{i=1}^2 \left[ \frac{p_x^{(i)2} + p_y^{(i)2}}{2m} - \frac{qB}{2m} x^{(i)} p_y^{(i)} + \frac{q^2 B^2}{4m} x^{(i)2} \right] \nonumber \\
&\quad - \frac{q^2 B^2}{4m^2 \Delta} x^{(1)} p_y^{(1)} x^{(2)} p_y^{(2)} - \frac{q^4 B^4}{16 m^2 \Delta} x^{(1)2} x^{(2)2}.
\nonumber  
\end{align}

One can see that the interaction terms treated as a perturbation of a single-particle Landau Hamiltonian  introduce nonlinearity and flux-mediated interactions between the particles. The first term in the second line describes angular momentum coupling, or a current--current-type interaction between the two particles via \( x^{(1)} p_y^{(1)} x^{(2)} p_y^{(2)} \).
A potential \( x^{(1)2} x^{(2)2} \) correlates the particles' positions along \( x \), favoring both particles being far from the center.
%maybe something intersting like fractional states
The interaction terms in $H_{\mathrm{eff}}$ involve products of operators associated with spatially separated particles, such as $x^{(1)} p_y^{(1)} x^{(2)} p_y^{(2)}$, and do not decay with the inter-particle distance. Unlike classical interactions (e.g., Coulomb, dipole-dipole), which are functions of $|\mathbf{r}_1 - \mathbf{r}_2|$, these flux-mediated terms arise from a shared coupling to a global quantum degree of freedom. As a result, the interaction acts regardless of spatial separation, establishing it as fundamentally nonlocal. For many particles, the interaction terms will be all-to-all.  

The $A^2$ term is included so that the model is gauge invariant. The interaction is attractive when the two particles have the same current direction. It will be interesting to explore if this interaction between the two particles exchanging a fluxon could result in tunable superconductivity.

In the Aharonov--Bohm (AB) effect, a charged particle acquires a geometric phase upon encircling a region of the magnetic flux, even if there is no magnetic field along the particle's path. In our system, a similar effect arises dynamically: when a particle encircles a localized quantized flux, the joint wavefunction of the particle and the flux qubit becomes  an entangled state of the form
\begin{equation}
    \alpha |\psi\rangle \otimes |0\rangle + \beta e^{i \theta} |\psi\rangle \otimes |1\rangle, 
    \nonumber 
\end{equation}
where $|\psi\rangle$ is the particle's wavefunction and $\theta = 2\pi \Phi/\Phi_0$ is the AB phase associated with the quantized flux state $|1\rangle$. This process results in a nontrivial Berry phase and entanglement between the matter and gauge field.

Our effective interaction emerges from integrating out a discrete gauge quantum DOF. 
In our system, the quantized flux qubit plays the role of a truncated gauge field, and the resulting second-order interaction is structurally similar to the effective interactions derived from the Dicke model of atoms coupled to a cavity mode.

%=================================================
\section{Many-particle system coupled to the QMF}
%=================================================

Now consider a 2D gas of $N$ charged particles (\(q = {-}e\), mass \(m\)) interacting  with a quantized
magnetic flux described by a two‑level system with energy gap~\(\Delta\).
Choosing the symmetric gauge  
\(\mathbf A = \frac{B}{2}(-y,x,0)\)  and  
\(\hat n=\tfrac12(1+\sigma^z)\) for the flux occupation, the
 exact Hamiltonian reads
\begin{equation}\label{eq:Hfull}
\hat H=\sum_{i=1}^{N}\frac{[\bm p_i-q\hat n\,\bm A(\bm r_i)]^{2}}{2m}
      +\Delta\,\sigma^{x}\; .
\end{equation}
This  case is harder to analyze analytically than the particles on a ring case. However, many properties of the system can be understood from studying the structure of different terms. 
Expanding the square gives
\begin{equation}
\hat H=\hat H_{\rm kin}+\tfrac12\hat F+\tfrac12\sigma^z\hat F
       +\Delta\,\sigma^{x},\;
\hat H_{\rm kin}=\sum_{i}\frac{\bm p_i^{2}}{2m},
\nonumber 
\end{equation}
with the \emph{collective operator}
\begin{equation}
\hat F=-\omega\sum_{i}\hat\lz^{(i)}
       +\frac{m\omega^{2}}{2}\sum_{i}\hat\rr_i,
       \nonumber 
\end{equation}
where for each particle $\lz=x p_y-y p_x$, $r^2=x^2+y^2$, 
and $\omega\equiv\frac{qB}{2m}$. In the first order approximation the effective $B$ is just the average of 0 and $B$.

Since the orbital and QMF DOFs couple through the term $\tfrac12\sigma^z\hat F$, it is interesting to analyze it. The operator $L_z$ commutes with the Hamiltonian. The coupling strength depends on $\omega$ so it can be in the ultrastrong coupling regime, since there is no restriction on the relative values of $\omega$ and $\Delta$. The total radius $\sum \hat{r}^2_i$ is coupled to the pseudospin, giving rise to collective dynamics.   
%-------------------------------------------------
\subsection{Elimination of the QMF}
%-------------------------------------------------

For \(\Delta \gg\omega \) the QMF degree of freedom can be again removed perturbatively via a
Schrieffer–Wolff transformation.
Keeping terms up to order \(1/\Delta^2\) one obtains the block‑diagonal
effective Hamiltonian
\begin{align}
\hat H_{\rm eff}
=\hat H_{\rm kin}
+\frac12\hat F
-\frac{1}{8\Delta}\hat F^{2}
+\frac{1}{128 \Delta^{3}}\hat F^{4}.
\label{eq:Heff_general}
\end{align}
Splitting  
\(\hat F=\hat F_1+\hat F_2\) with  
\(\hat F_1=-\omega\sum_i L_{iz} \) and
\(\hat F_2=(m\omega^{2}/2)\sum_i\rr_i\), one can see that the quadratic term contains an all to all angular momentum coupling, the terms that renormalize single-particle energies, and the terms describing the coupling of spatial radii and angular momenta term.

%-------------------------------------------------
\subsection{ Quantum phase transition}
%-------------------------------------------------
%As we discussed in the simpler ring case, the 2D gas case will exhibit phase transition for critical values of $\phi$ or $N$ 
As in the 1D case, phase transition in the thermodynamic limit will require either extensive $\Delta$ that scales with N, or scaling the coupling $\omega$ properly with $N$,  or relaxing the two level approximation of the QMF to allow excitation to higher states, taking into account  full energy spectrum of a linear LC resonator or a nonlinear resonator (superconducting flux qubit).

Next, we investigate the interactions due to nonlinear in $F$ terms in the Hamiltonian \eqref{eq:Heff_general} and whether they could result in the phase transition in a mean field theory. 
If one introduces the anti-confining potential $-\kappa r^2$, at a critical value of $\kappa$  it will destabilize the Landau levels, closing the gap of the system. However, due to the nonlinear in $F$ terms the system will find the stable state and will exhibit zero temperature phase transition with the  order parameter $X$ describing the size of the system, as defined below.

As the Hamiltonian commutes with all $L_z$, for a fixed $L_z$ one can write 
\begin{multline}
H=\sum_{i=1}^N\frac{\mathbf p_i^2}{2m}
+\frac{\kappa}{2}X
+\frac12F-\frac{1}{8\Delta}F^2+\frac{1}{128\Delta^3}F^4,\\ F=-\omega L+\frac{m\omega^2}{2}X,
\nonumber 
\end{multline}
where $X=\sum_i r_i^2$ and $L=\sum_i L_{iz}$. In a fixed-$L$ sector the minimal kinetic
energy at a given $X$ is $E_{\rm kin}^{\min}=L^2/(2mX)$, so the energy at $T = 0$  as a function of $X$ is (up to a constant) 
\begin{equation}
\mathcal E(X|L)=\frac{L^2}{2mX}+A\,X+B\,X^2+C\,X^3+D\,X^4,
\label{eq:ELandau}
\end{equation}
where expanding $F$ gives
\begin{align}
A&=\frac{\kappa}{2}+\frac{m\omega^{2}}{4}+\frac{m\omega^{3}L}{8\Delta}-\frac{m\omega^{5}L^{3}}{64\Delta^{3}}, 
 \;
C =-\frac{m^{3}\omega^{7}L}{256\Delta^{3}}, \nonumber\\
B&=-\frac{m^{2}\omega^{4}}{32\Delta}+\frac{3m^{2}\omega^{6}L^{2}}{256\Delta^{3}}, 
 \; 
D =\frac{m^{4}\omega^{8}}{2048\Delta^{3}}>0.
\label{eq:ABCD}
\end{align}

Eliminating the cubic term by $X=Y-\frac{C}{4D}$,  the polynomial part becomes a tilted quartic: $$\mathcal E_{\rm poly}(Y)=D\,Y^4+\tilde B\,Y^2+\tilde A\,Y,$$
where 
\begin{equation}
\tilde B=B-\frac{3C^2}{8D},\quad
\tilde A=A-\frac{BC}{2D}+\frac{C^3}{8D^2}.
\nonumber 
\end{equation}

\begin{figure}
    \centering
    \includegraphics[width=.99\linewidth]{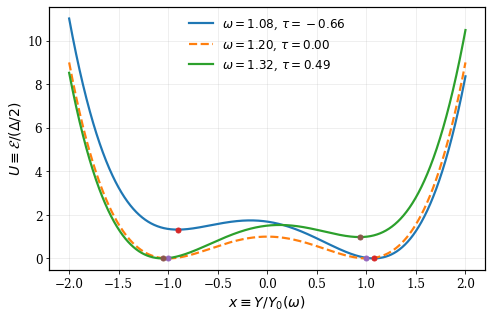}
    \caption{Universal tilted quartic size potential $U(x)=x^{4}-2x^{2}+\tau x$ plotted versus the dimensionless coordinate $x=Y/Y_{0}(\omega)$, with $U=\mathcal{E}/(\Delta/2)$ and $Y_{0}(\omega)=4\sqrt{2}\,\Delta/(m\omega^{2})$. Curves correspond to $\omega=\{0.90,1.00,1.10\}\,\omega_{t}$ at fixed $\kappa=-0.72$ and $m=1$ and $\Delta=1$, where $\omega_{t}=\sqrt{-2\kappa/m}=1.20$. Dots mark the local minima; each curve is vertically offset by subtracting its minimum. The control tilt is $\tau(\omega)=\tfrac{4\sqrt{2}}{m\omega^{2}}\,[\kappa-\kappa_{c}(\omega)]$ with $\kappa_{c}(\omega)=-m\omega^{2}/2$; thus $\tau<0$ favors the compact (left) well, $\tau=0$ gives coexistence, and $\tau>0$ favors the swollen (right) well.}
    \label{fig6}
\end{figure}

A direct substitution using Eq.~\eqref{eq:ABCD} yields the exact simplifications
(independent of $L$ for $\tilde B$ and of $\Delta$ for $\tilde A$):
\begin{equation}
{\;\tilde B(\omega)=-\frac{m^2\omega^4}{32\Delta}<0,\qquad
\tilde A(\omega)=\frac{\kappa}{2}+\frac{m\omega^2}{4}.}
%\label{eq:controls_simple}
\nonumber 
\end{equation}
Thus the quartic is intrinsically a double-well potential ($\tilde B<0$), as shown in Fig.~\ref{fig6}; the sign of $\tilde A$
tilts it and changes at the “untilt” point
\begin{equation}
\tilde A(\omega_t)=0\ \Rightarrow\ 
{\ \omega_t^2=\frac{-2\kappa}{m}\quad(\kappa<0).}
\nonumber 
\end{equation}

The boundaries of metastability (spinodals) follow from the cusp condition of a tilted
quartic,
\begin{equation}
8\,\tilde B^3+27\,D\,\tilde A^2=0,
\qquad D(\omega)=\frac{m^4\omega^8}{2048\Delta^3},
\nonumber 
\end{equation}
which reduces to
$m^2\omega^4=54\big(\tfrac{\kappa}{2}+\tfrac{m\omega^2}{4}\big)^2$ and yields two
spinodal frequencies $\omega_{\rm sp,1}<\omega_{\rm sp,2}$:
\begin{equation}
{\;
\omega_{\rm sp,1}^{2}
\approx 1.3\,\frac{-\kappa}{m}, \;
\omega_{\rm sp,2}^{2}
\approx 4.4\,\frac{-\kappa}{m}.}
\end{equation}
For $\omega_{\rm sp,1} < \omega < \omega_{\rm sp,2}$ the potential has two minima; along an
inner coexistence line they are degenerate and the equilibrium value $X_\ast$ has a jump 
(the first-order transition). In the symmetric limit (ignoring the small-$X$ kinetic
bias) the minima sit at $Y=\pm Y_0$ with
\begin{equation}
Y_0=\sqrt{-\frac{\tilde B}{2D}}
=\frac{4\sqrt{2}\,\Delta}{m\,\omega^2}\quad\Rightarrow\quad
\Delta X\simeq 2Y_0,
\end{equation}
evaluated near $\omega\simeq\omega_t$; the centrifugal term $L^2/(2mX)$ shifts the
coexistence slightly to $\omega_{\rm coex}(L)\gtrsim\omega_t$ but does not alter the
spinodals. A continuous ($2^\text{nd}$-order) point would require $\tilde A=\tilde B=0$,
which is impossible for $\omega>0$; hence the transition is generically
first order for $\kappa<0$.

For fermions at $T = 0$ the Pauli pressure adds a small-$X$ barrier
$\gamma/X$ (with $\gamma>0$) to the size energy. This term \emph{does not} modify
$\tilde A,\tilde B,$ or $D$, and therefore leaves the \emph{topology} of the phase
diagram (spinodals, first-order character, absence of a second-order point) intact.
Quantitatively it lifts the compact minimum and thus shifts the coexistence toward
weaker anti-confinement,
\begin{equation}
\omega_{\rm coex}^{\rm(ferm)}(L)\lesssim \omega_{\rm coex}^{\rm(bare)}(L),
\nonumber 
\end{equation}
and typically increases the jump $\Delta X$ at coexistence. For finite $N$ the
singularity is rounded into a narrow avoided crossing; the first-order behavior
sharpens with $N$ and becomes exact in the mean-field limit.

The small-$X$ phase is compact (high density, larger $\Omega=L/mX$),  while the large-$X$ phase is swollen (dilute core, smaller $\Omega$).  Without confinement ($u=\kappa=0$) no true quantum phase transition occurs (only crossover/metastability). 
The nonlinearity coming from the $F^2$ term and higher terms will affect the edge states velocity. 

The realization of such system can happen with an artificial gauge field, e.g., in the case of cold atoms where an effective anti-confining potential can be realized, or with electrostatic gating of quantum Hall systems; of course the system should be confining at large radii $R$.

\section{Opposite direction states of QMF}

In the discussion above we focused on the case where the magnetic field is described by the operator $\hat{n} B$ representing coupling between free particles and Landau-quantized states.  It describes the hybridization of free and Landau-quantized electrons, or two topologically distinct states \cite{hybridtopology}.  However, there is also the possibility of having the magnetic field as a superposition of the kind $\hat{I}B_{ext}+\hat{n} B_0$ where $B_{ext}$ is a classical field bias and $\hat{I}$ is a 2x2 unit matrix. Such a system will have different properties because it describes the hybridization of two Landau-quantized systems. Time reversal symmetry is completely broken in this case. An external classical flux can shift the QMF values so that the two states $\hat{n}(2B_0)- B_0$ correspond to same magnitude but opposite directions.

To begin with, we consider a one-particle problem in a uniform perpendicular field $\vb B = B\,\hat{\vb z}$ in the Landau gauge, when  
\begin{equation}
\hat {H}={\frac {{\hat {p}}_{x}^{2}}{2m}}+{\frac {1}{2m}}\left({\hat {p}}_{y}-q(IB_{ext}+\hat{n} B_0){\hat {x}}\right)^{2}+\Delta\sigma^{x}. 
\label{6-1}
\end{equation}

\begin{figure}[t]               % placement “t” for top; use h or b if preferred
  \centering
  %--- sub-figure (a) ---
  \subfloat[]{%
    \includegraphics[width=0.50\linewidth]{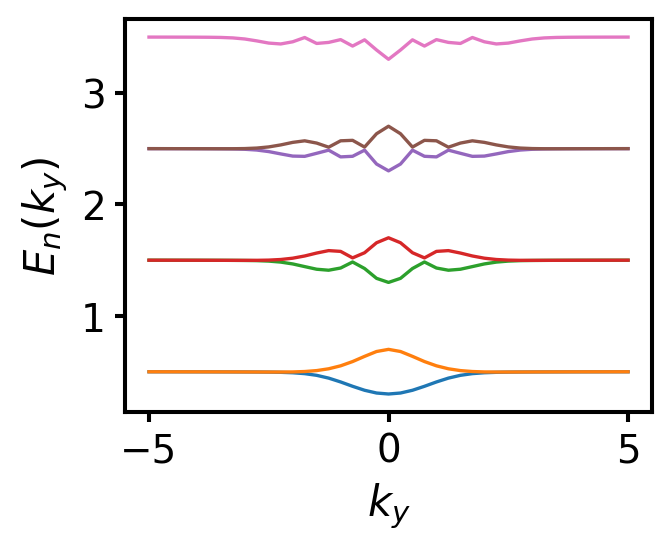}%
    \label{fig:panel_a}}%
  \hfill
  %--- sub-figure (b) ---
  \subfloat[]{%
    \includegraphics[width=0.45\linewidth]{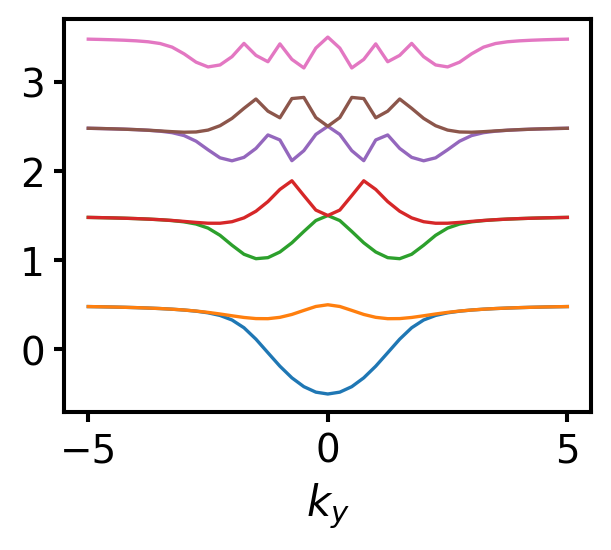}%
    \label{fig:panel_b}}%
  \hfill
  %--- sub-figure (c) ---
   \subfloat[]{%
    \includegraphics[width=0.450\linewidth]{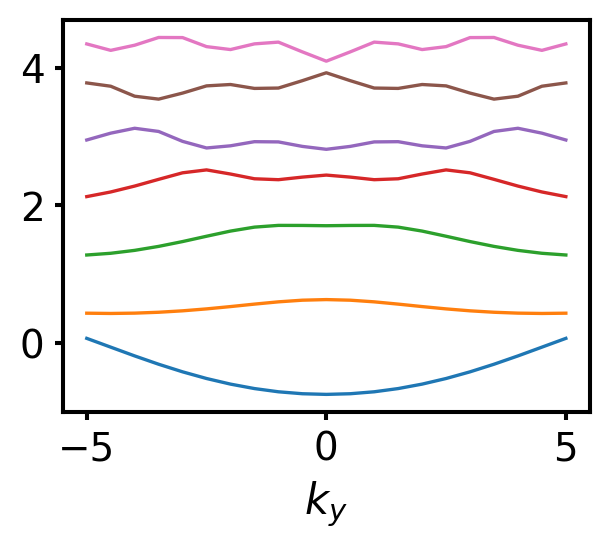}%
    \label{fig:panel_b}}%
     \hfill
%---------------D----------
   \subfloat[]{%
    \includegraphics[width=0.450\linewidth]{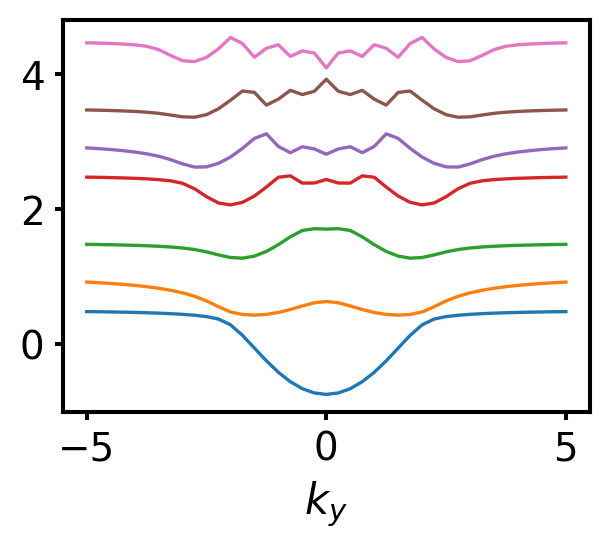}%
    \label{fig:panel_b}}%
     \hfill

  \caption{    
Eigenenergies in the Landau gauge for the Hamiltonian Eq.~(\ref{6-1}). (a) $\Delta=1$, (b) $\Delta=0.2$, (c) $\Delta=1.5$, and the fields are $B = \pm 1$ for two Landau-quantized systems. (d) $\Delta=1.5$ with asymmetric superposition of $B=-2$ and $B=1$. All energies are normalized by $\omega_c$ and the fields are normalized by some arbitrary value. 
   }
    
  \label{fig:three_panel}
\end{figure}

One can diagonalize this Hamiltonian numerically. The resulting eigenenergies are shown in  Fig.~\ref{fig:three_panel}. 
The parts (a) to (c) show the hybridization of the two degenerate energy ladders corresponding to the two equal-value and opposite-direction fields $\pm B$ and different values of $\Delta$.
The effect of the QMF is pronounced at small values of $k_y$. As $\Delta$ increases, the influence extends to larger $k_y$. For a finite sample, this modification will impact the edge states. One example of the asymmetric bias field effect is shown in Fig.~\ref{fig:three_panel}(d), illustrating the mixing of two Landau-quantized systems.

\subsection{Exact diagonalization in the symmetric gauge}

It turns out that an exact analytic diagonalization of this Hamiltonian is possible in the symmetric gauge
$\vb A(\vb r)=\tfrac{B}{2}(-y,x,0) = \tfrac{1}{2}\,\vb B\times\vb r$, even in the many-particle case, which allows us to explore quantum phase transitions.  Using the exact identity
\begin{equation}
\vb A\!\cdot\!\vb p = \frac{B}{2}\,L_z,\; 
L_z \equiv x p_y - y p_x ,
\nonumber 
\end{equation}
the Hamiltonian 
\begin{equation}
 \hat{H}=\sum_{i=1}^{N}\frac{\big[\vb p_i - q\,\sigma_z\,\vb A(\vb r_i)\big]^2}{2m}
+\Delta\,\sigma^{x}
\nonumber 
\end{equation}
can be written as
\begin{align}
 H
&=\sum_{i=1}^{N}\!\left[\frac{\vb p_i^{\,2}}{2m}
+\frac{q^{2}B^{2}}{8m}\,r_i^{2}\right]
\;-\;\frac{\omega_c}{2}\,\sigma_z\,\sum_{i=1}^{N} L_{z,i}
\;+\;\Delta\,\sigma^{x}
%\label{eq:H_matrix1}
\nonumber 
\\[4pt]
&=  K+\Lambda\,\sigma^{z}+\Delta\,\sigma^{x},
\label{eq:H_matrix2}
\end{align}
where 
\begin{equation}
K=\sum_{i=1}^N\!\left[\frac{\vb p_i^{\,2}}{2m}+\frac{q^2B^2}{8m}\,r_i^2\right],\; \Lambda = -\,\frac{\omega_c}{2}\,L,\; 
L = \sum_{i=1}^{N} L_{z,i}.
\label{kdef} 
\end{equation} 

Integrating out the QMF in the adiabatic limit yields, to leading nontrivial order, to a particles-only Hamiltonian $$ H_{\rm eff}=H_{\rm e}-\Delta- \frac{\omega_c^{2}}{8\Delta} \big(\sum_{i}\hat L_{z,i}\big)^{2}+O(\Delta^{-3})$$
with $$ H_{\rm e}=\sum_{i}\left[ \frac{\mathbf p_i^{2}}{2m}+\frac{q^{2}B^{2}r_i^{2}}{8m}\right]. $$  
The induced term 
$$ -\frac{\omega_c^{2}}{8\Delta} \big(\sum_{i}\hat L_{z,i}\big)^{2}= - \frac{\omega_c^{2}}{8\Delta} \,\left[ \sum_{i}\hat L_{z,i}^{2}+2\sum_{i<j}\hat L_{z,i}\hat L_{z,j} \right] $$ 
is unusual in that it is (i) all-to-all and distance-independent. This symmetric configuration of the two states of the field, unlike the one in Sec.~IV A, does not have any spatial dependence; (ii) attractive in the orbital (chiral) channel, favoring aligned angular momenta and hence large \(|K|\equiv\big|\sum_i \hat L_{z,i}\big|\) (iii) diagonal in the \(|m\rangle\) basis with pairwise weight \(\propto m_im_j\), unlike short-range or Coulomb matrix elements, (iv) tunable by varying the magnetic field and the gap via the term \(\propto B^{2}/\Delta\). While the statistics has not been taken into account yet, we expect this interaction to result in an orbital Stoner tendency for fermions, thereby enabling spontaneous time-reversal breaking once its gain exceeds single-particle costs.  Note the requirement \((\omega_c/2\Delta)\,\|K\|\ll 1\), with next corrections entering at \(O(\Delta^{-3})\).

There is a symmetry in the coupling term, corresponding to flipping the direction of $L$ and the QMF state, which makes  analytic diagonalization possible and has profound physical consequences. 
Rotational invariance implies $[K,L]=0$, hence $[K,\Lambda]=0$. Therefore,
in any common eigen-sector of $(K,L)$ the qubit ``sees'' a $2\times2$ matrix 
\begin{equation}
h(\Lambda)=\Lambda\,\sigma^{z}+\Delta\,\sigma^{x} 
\nonumber 
\end{equation}
with eigenvalues $\pm R$, where
\begin{equation}
R=\sqrt{\Delta^{2}+\Lambda^{2}}.
\nonumber 
\end{equation}
We see that the total angular momentum of particles is giving rise to an effective z-field component of the pseudospin. The lower qubit branch yields the orbital  Hamiltonian 
\begin{equation}
H
= K-\sqrt{\Delta^{2}+\Lambda^{2}}\; .
%\label{eq:Heff_exact}
\nonumber 
\end{equation}
The qubit eigenvectors depend only on $L$,
which commutes with $K$. 
%In terms of the total angular momentum $L$ only.
Note that the operator $K$ in Eq.~\eqref{kdef} describes 
 a 2D isotropic oscillator with frequency $\omega_c/2$ where $\omega_c=\frac{qB}{m}$. Hence 
\[
{\,E_{n_r,m}= \frac{\omega_c}{2}\big(2n_r+|m|+1\big),\; n_r=0,1,2,\dots,\; m\in\mathbb Z.}
\]
So the total energy is 
\[{\,E=\sum_{n_r,m} \frac{\omega_c}{2}\big(2n_r+|m|+1\big)-\sqrt{\Delta^{2}+(\sum_m\frac{\omega_c}{2} m}}
)^{2}. 
\]
%The ground state is always $n_r=0,m=0$.
In the usual setup, the allowed values of $m$ (or chirality) depend on the direction of $B$: 
for either $\pm B$, the states with $\pm |m|$ are allowed. 
However, in the present case the $\pm m$ states are degenerate.

\subsection{Quantum phase transition for fermions}

We consider $N$ fermions occupying distinct angular-momentum orbitals $m\in\mathbb{Z}$ in the $n_r=0$ sector.
The total energy for a configuration $\{m_i\}_{i=1}^N$ is
\begin{equation}
E(\{m_i\})
=\frac{\omega_c}{2}\Big(\sum_{i=1}^N |m_i| + N\Big)
-\sqrt{\Delta^2 + \Big(\frac{\omega_c}{2}\sum_{i=1}^N m_i\Big)^2}\,.
\label{eq:Edef}
\end{equation}
Introduce
\begin{equation}
S \equiv \sum_{i=1}^N |m_i|,\qquad
M \equiv \sum_{i=1}^N m_i,\qquad
\omega \equiv \frac{\omega_c}{2},
\end{equation}
so that
\begin{equation}
E = \omega (S+N) - \sqrt{\Delta^2 + (\omega M)^2}.
\label{eq:Ealpha}
\end{equation}
The first term penalizes large $|m|$ individually, while the second (negative) term rewards large $|M|$ collectively.
Given the Pauli principle, the problem reduces to an integer optimization in the set $\{m_i\}$.
Two extremal patterns compete to dominate:
%and bracket all possibilities:

\paragraph*{(i) Balanced set, $M=0$.}
Choose occupied $m$ values symmetrically about zero so as to minimize $S$:
\begin{align}
N=2k\ \text{(even)}: \quad & m=\pm 1,\pm 2,\dots,\pm k, \nonumber\\
& S_{\rm bal}=k(k+1)=\frac{N^2}{4}+\frac{N}{2}; \\
N=2k+1\ \text{(odd)}:\quad & m=0,\pm 1,\dots,\pm k, \nonumber\\
& S_{\rm bal}=k(k+1)=\frac{N^2-1}{4}.
\end{align}
So $ M_{\rm bal}=0$ and the energy is
\begin{equation}
E_{\rm bal}=\omega(S_{\rm bal}+N)-|\Delta|.
\label{eq:Ebal}
\end{equation}

\paragraph*{(ii) Fully polarized set, maximal $|M|$.}
Take consecutive nonnegative (or nonpositive) integers:
\begin{equation}
m=\{0,1,2,\dots,N-1\}\quad \text{or}\quad \{0,-1,-2,\dots,1-N\}.
\nonumber
\end{equation}
Then $$S_{\rm pol}=M_{\rm pol}=\frac{N(N-1)}{2},$$
and 
\begin{equation}
E_{\rm pol}=\omega\!\left(\frac{N(N-1)}{2}+N\right)
-\sqrt{\Delta^2 + \big(\omega \tfrac{N(N-1)}{2}\big)^2}.
\label{eq:Epol}
\end{equation}

For a fixed $N$, the ground state is either balanced or fully polarized. Indeed, any partially polarized set increases $S$ relative to the balanced choice without achieving a larger $|M|$ than the fully polarized one; since $E$ is increasing in $S$ (linearly) and decreasing in $|M|$ (concavely), no intermediate pattern can beat both extrema.

\subsection{Phase boundary $\Delta_c$ and ground state}

Equating $E_{\rm pol}=E_{\rm bal}$ yields a closed form for the critical coupling condition which separates the two phases.
Let
\begin{equation}
a \equiv \omega\,M_{\rm pol}=\omega\,\frac{N(N-1)}{2},\; 
y \equiv \omega\,(S_{\rm pol}-S_{\rm bal}),
\nonumber 
\end{equation}
so that $E_{\rm pol}=E_{\rm bal}$ becomes
\begin{equation}
y = \sqrt{\Delta_c^2+a^2}-\Delta_c
\; \Longrightarrow \;
{\ \Delta_c=\frac{a^2-y^2}{2y}\ }\qquad (y>0).
\label{eq:Dc_general}
\end{equation}
Evaluating the difference $S_{\rm pol}-S_{\rm bal}$ gives two cases.

\paragraph*{Odd $N$.}
Here $S_{\rm pol}-S_{\rm bal}=\frac{(N-1)^2}{4}$ and $y=\omega\frac{(N-1)^2}{4}>0$, hence
\begin{equation}
 \Delta_c  = \frac{\omega_c}{16}\,(3N^2+2N-1)\,. 
\label{eq:Dc_odd}
\end{equation}

\paragraph*{Even $N$.}
Here $S_{\rm pol}-S_{\rm bal}=\frac{N(N-4)}{4}$ and $y=\omega\frac{N(N-4)}{4}$.
For $N\le 4$, $y\le 0$, so $E_{\rm pol}\le E_{\rm bal}$ \emph{for all} $\Delta\ge 0$ (the polarized phase always wins).
For $N>4$ ($y>0$),
\begin{equation}
\Delta_c =\frac{\omega_c}{16}\,\frac{3N\,(N^2-4)}{(N-4)}\,. 
\label{eq:Dc_even}
\end{equation}

One can think of this transition as competition between the momentum space attraction due to the coupling with the global QMF and the Pauli repulsion for similar momentum states. Including the spin of the fermions will modify the numerical factor in front of $\omega_c$ but will keep the $N^2$ scaling.

\subsubsection{Ground-state selection}

To summarize the above options, the ground state is 
\begin{equation}
%\label{eq:GS_rule}
\begin{aligned}
\mathrm{GS}=
&\begin{cases}
\text{balanced }(M=0), & \Delta>\Delta_c,\\
\text{fully polarized }(|M|=\tfrac{N(N-1)}{2}), & \Delta<\Delta_c,
\end{cases} \\[-2pt]
\end{aligned}
\nonumber 
\end{equation}
where for odd $N$ one should use Eq.~\eqref{eq:Dc_odd} and for 
even $N>4$ one should use Eq.~\eqref{eq:Dc_even}. 

For even $N\le4$ the GS is polarized for all $\Delta$. In the polarized phase the ground state is twofold degenerate (all $m\ge 0$ or all $m\le 0$), since the energy $E$ depends on $M^2$. 

In the experimental setting it might be more feasible to vary other parameters rather than $\Delta$. Depending on which parameters are fixed and which can be varied, the above expressions for $\Delta_c$ can be used for fixed $\Delta$ to find critical values of $N$ or $\omega_c$. 

In the balanced phase two edge channels with opposite chirality exist simultaneously. while in the polarized phase the system will be in one of the two channels.

For the balanced case the qubit stays in the ground state of $\sigma_x$ so the particles have no effect on it, while in the polarized case the total $M$ acts as an effective z-field that shifts the state of the qubit towards the poles on the Bloch sphere. In this latter case the state of the QMF is not entangled with the state of the particles. There is a certain analogy  with the Stoner model of ferromagnetism.

\begin{equation}
E_{\rm bal}=
\begin{cases}
\omega\!\left(\dfrac{N^2}{4}+\dfrac{3N}{2}\right)-|\Delta|, & N\ \text{even},\\[8pt]
\omega\!\left(\dfrac{N^2-1}{4}+N\right)-|\Delta|, & N\ \text{odd},
\end{cases}
\nonumber 
\end{equation}
\\ 
\text{and} 
\begin{equation}
 E_{\rm pol}=\omega\!\left(\dfrac{N(N-1)}{2}+N\right)
-\sqrt{\Delta^2+\big(\omega \tfrac{N(N-1)}{2}\big)^2}.
\nonumber 
\end{equation}

\noindent
These formulas make the competition between positive ($S$) and negative energy terms (via $|M|$) explicit and quantify the balanced to polarized phase crossover as a sharp threshold in $N$ at fixed  $\Delta$.

%%%%%%%%%%%%%%%%%%%%%%%%%%%%%%%%%%%%%%%

\subsection{Bosons instability and phase transition}

In the case of bosonic particles, any finite deconfinement potential will cause a transition. Indeed, 
 consider $N$ spinless bosons confined to a 2D plane. We introduce the rotationally symmetric deconfining potential 
\begin{equation}
V_{\rm conf}(r)=-\tfrac12 m\omega_0^2 r^2 
\nonumber.
\end{equation}
The orbital motion sector is the Fock--Darwin problem.  Using the commuting pair of operators $L_z$ and 
\begin{equation}
H_0=\frac{\vb p^2}{2m}+\frac{m\omega_c^2 r^2}{8}+\tfrac12 m\omega_0^2 r^2,
\nonumber 
\end{equation}
we work in the joint eigenbasis $\ket{n_r,m}$ (one again needs to distinguish between the mass $m$ and the $L_z$ eigenvalue $m$):
\begin{equation}
H_0\ket{n_r,m}=E_0(n_r,m)\ket{n_r,m},\; 
L_z\ket{n_r,m}=\hbar m\ket{n_r,m},
\nonumber
\end{equation}
where 
\begin{equation}
E_0(n_r,m)=\frac{\Omega}{2}\,(2n_r+|m|+1), \; \Omega \equiv 2\sqrt{\frac{\omega_c^2}{4}-\omega_0^2}.
\label{34} 
\end{equation}
The branch relevant to the ground state is
\begin{equation}
E_-= \sum_{n_r,m}\frac{\hbar\Omega}{2}(2n_r\!+\!|m|\!+\!1)
-\sqrt{\Delta^2+\Big(\frac{\hbar\omega_c}{2}M\Big)^2}.
\nonumber 
%\label{eq:Eminus}
\end{equation}

If  $\Omega\ge |\omega_c|$, then %$f'(S)\ge 0$ for all $S\ge 0$.
the minimum energy is realized by putting \emph{all} bosons in the $\ket{n_r\!=\!0,m\!=\!0}$ state, so that $M=0$ and
\begin{equation}
E_{\rm GS}=\frac{\omega'}{2} N-|\Delta|=\frac{\hbar\Omega}{2}\,N-|\Delta|.
\nonumber 
\end{equation}
%,\ket{\Psi_{\rm GS}}=\frac{(\hat a^\dagger_{0,0})^N}{\sqrt{N!}}\ket{0}\otimes\frac{\ket{\downarrow}-\ket{\uparrow}}{\sqrt2}.
 If $\Omega<|\omega_c|$, then %sufficiently large $S$ one has $f'(S)<0$ and $f(S)$ decreases with $S$; so 
 the model exhibits a runaway tendency (ill-posed in the absence of a boundary or cutoff). In any realistic finite sample with a maximal accessible $|m|$, the minimum occurs at the boundary $S^*=S_{\max}$: \emph{an edge-polarized phase}.

What is interesting here is that any finite $\omega_0$ will make the system unstable and all the bosons will fall to the maximum allowed $m$ state, essentially to the edge. In the absence of $\Delta$ this is not the case as one would need $\omega_0>\omega_c$ for instability. The collective coupling renders the system sensitive to destabilization making it a platform for studying classical and quantum chaos. 

\subsection{ Remark on hard walls}

For a hard-wall disk (finite radius $R$) the single-particle energies develop a positive edge dispersion $\Delta E_{\rm edge}(m)$ as the guiding center approaches the boundary; the cost of pushing a boson to the outermost allowed $|m|$ is finite, $\Delta E_{\rm edge}=O(\hbar|\omega_c|)$, while the allowed $|m|$ extend up to $m_{\max}\sim R^2/(2\ell_B^2)$ with $\ell_B=\sqrt{\hbar/(m|\omega_c|)}$. 
In this case the competition is between a \emph{bounded} edge cost (per boson) and a \emph{linear} qubit gain $-\sqrt{\Delta^2+\omega_c^2 S^2}\sim -\omega_c S$ for large $S$.
Consequently, for sufficiently large samples one finds an edge-polarized ground state with $S^*=S_{\max}=N m_{\max}$ whenever
\begin{equation}
\omega_{crit}\,m_{\max} \;>\; \Delta E_{\rm edge},
\nonumber 
\end{equation}
i.e., the size/field beyond a threshold favors maximal same-sign $m$ occupation. 
Thus, with hard walls the qualitative ``bulk ($S=0$) vs edge-polarized ($S=S_{\max}$)'' first-order transition persists, with the role of $\omega'$ replaced by the finite edge penalty $\Delta E_{\rm edge}$ and the geometric cutoff $m_{\max}$.

\section{Tight binding model and artificial gauge fields}

Neutral particles, such as photons or ultracold atoms in artificial gauge fields, behave as though influenced by EM fields, mimicking effects that normally require charged particles. These synthetic fields enable tabletop simulations of fascinating and complex phenomena, from quantum Hall effect and Hofstadter's butterfly to lattice gauge theories and topological insulators \cite{Dalibard2011,CooperRMP2019,Ozawa2019,Aidelsburger2013,Miyake2013,Banuls2020}. Photonic platforms provide a versatile setting for engineering artificial gauge fields. By tailoring lattice geometries, temporal modulations, or wavefront properties, synthetic gauge potentials for light can be realized \cite{Ozawa2019,Hafezi2013,Fang2012,Longhi2013,Lin2015}. Helical waveguide arrays implement synthetic magnetic flux through spatially varying propagation phases, enabling the observation of photonic Floquet topological insulators \cite{Rechtsman2013,Ozawa2019}. Orbital angular momentum (OAM)–engineered beams injected into photonic lattices can induce tunable synthetic flux, leading to effects such as Aharonov–Bohm caging \cite{ABcage}.

Dynamic modulation of resonators or coupled waveguides creates effective electric and magnetic fields for photons, realizing robust edge transport and nonreciprocal light flow \cite{modulation}. 

For particles that are naturally noninteracting, for example photons, our scheme aims to induce interaction in the topological systems where artificial gauge fields are quantized and the interaction is tunable and sensitive to the DOF mediating the interaction. 

We start with the regular tight binding Hamiltonian and then introduce the Peierls phase as a quantum operator to imitate the QMF.
 Consider a 1D lattice with periodic boundary conditions (ring) and the hopping parameter $t$: 
$$H = -t \sum_{i} \left( e^{i \theta_{j}} a_j^\dagger a_{j+1} + \text{H.c.} \right)
$$
Here the variables $\theta_j$ describe a synthetic gauge field.

Without a synthetic field, 
    $
    H_0 = \sum_{k} \left( -2t \cos(k) \right) \hat{b}_k^\dagger \hat{b}_k $, whereas with uniform synthetic field $\theta$,  
    %{\bf what happened to subscript $\theta_j$ ? }
    $
    H_1 = \sum_{k} \left( -2t \cos\left(k + \theta\right) \right) \hat{b}_k^\dagger \hat{b}_k$,
where $\hat{b}_k$ is the annihilation operator for quasi-momentum $k$. 
    %{\bf define $b$  operators}
    
Now we upgrade $\theta$ to an operator $\hat{n}\theta$, where the average $n$ can be any real number since this is an engineered synthetic flux. Then the Hamiltonian becomes  
\[
    H = \sum_{k} \left( -2t \cos\left(k + \hat{n}\theta\right) \right) \hat{b}_k^\dagger \hat{b}_k + \Delta \sigma_{x}
    \]
    or in the matrix form,
\begin{align} 
\begin{bmatrix}
\sum_{k} \left( -2t \cos\left(k + n_{1}\theta\right) \right) \hat{b}_k^\dagger \hat{b}_k & \Delta \\ \Delta & \sum_{k} \left( -2t \cos\left(k +n_{2}\theta\right) \right) \hat{b}_k^\dagger \hat{b}_k
\end{bmatrix}
\label{eq:f21}
\end{align}

After setting the hopping parameter $t=1$ and $\theta=\pi/2$, the energy for N bosons with quasi-momentum $k$ is given by 
$$
\frac{1}{2} \left( -N (\cos(k)+ \sin(k)) \pm \sqrt{4 \Delta^2 + N^2 - N^2 \sin(2k)} \right)
$$

We notice strong nonlinear dependence on $N$ and sensitivity to the value of $\theta$ (the flux). The analysis above can describe the case of photons in a synthetic QMF and its possible implementations will be discussed later. In this case, the single particle band structure has a band gap, i.e., the system is not transparent at certain frequencies; however, the existence of one photon closes the band gap making the system transparent. This is expected given the nonlinearity and can be used to entangle different photons, i.e., via a controlled photon blockade. Moreover, the emergent photon-photon interaction can be used to simulate the dynamics of interacting bosons on a lattice. The spectrum of the system is sensitive to small variations in the value of the QMF. This could enable observable interactions between single photons.  

Photon blockade phenomena can also be realized in atom-cavity systems. An  
advantage that we have here is sensitivity to variation in the macroscopic parameter of the system, avoiding the need to control the state of an individual atom. 

%small theta  not an atom ie macroscopic system tunabilty of all the parameters 

\subsection{Many particles on the ring in dispersive limit}

We start from the two-branch (qubit) single-particle dispersions
\begin{equation}
\varepsilon_j(k) = -2t\,\cos\!\big(k+n_j\theta\big),\qquad j=1,2,
\end{equation}
with a tunneling $\Delta$ between the branches. Setting $t=1$ and defining
\begin{equation}
\bar n=\frac{n_1+n_2}{2},\quad \delta n=n_1-n_2,\quad
\phi=\bar n\,\theta,
\nonumber 
\end{equation}
the $2\times 2$ block can be written as
\begin{equation}
H = \sum_k \bar\varepsilon(k)\,\hat n_k\otimes \mathbb{I}
+ g\,\hat J\otimes \sigma_z + \Delta\,\mathbb{I}\otimes\sigma_x,
\nonumber 
%\label{eq:H-reduced}
\end{equation}
where
\begin{align}
\bar\varepsilon(k) &= -2C\,\cos(k+\phi),\\
\hat J &= \sum_k 2\sin(k+\phi)\,\hat n_k,
\nonumber
\end{align}
and we introduced the geometric coefficients
\begin{equation}
C\equiv \cos\!\Big(\frac{\delta n\,\theta}{2}\Big),\quad
g\equiv \sin\!\Big(\frac{\delta n\,\theta}{2}\Big).
\nonumber 
\end{equation}
Here $\hat n_k=\hat b_k^\dagger \hat b_k$ is the occupation of a state with momentum $k$. Note that $[\hat n_k,\hat n_{k'}]=0$ and $[\hat n_k,H]=0$, so $\hat J$ is a function of the integrals of motion.

In the dispersive regime $|\Delta|\gg |g\hat J|$ the qubit remains near an eigenstate of $\sigma_x$. For a fixed many-body configuration (thus fixed $J$, the eigenvalue of $\hat J$), the qubit Hamiltonian is
\begin{equation}
H_q(J)= gJ\sigma_z + \Delta\,\sigma_x,
\nonumber 
\end{equation}
with exact eigenvalues $\pm \Omega(J)$,
\begin{equation}
\Omega(J)=\sqrt{\Delta^2 + g^2 J^2}.
\nonumber 
\end{equation}
Projecting onto the qubit ground band and expanding for large $|\Delta|$ gives the qubit ground-state energy
\begin{equation}
-\Omega(J)= -\Delta - \frac{g^2}{2\Delta}\,J^2 + \mathcal{O}\!\left(\frac{g^4 J^4}{\Delta^3}\right).
\nonumber 
\end{equation}
Dropping the constant $-\Delta$, the many-body effective Hamiltonian (to leading nontrivial order) is
\begin{equation}
{\,H_{\mathrm{eff}}=\sum_k \bar\varepsilon(k)\,\hat n_k
\;-\;\frac{g^2}{2\Delta}\,\hat J^{\,2}
\;+\;\mathcal{O}\!\left(\frac{g^4 \hat J^4}{\Delta^3}\right).\,}
\label{eq:Heff}
\end{equation}
Thus, adiabatic elimination mediates an all-to-all interaction \emph{in momentum space}, quadratic in the total current $\hat J$.

%%%%%%%%%%%%%%%%%%%%%%%%%%%%%%%%%%%%%%%%%%%

\subsection{Bosonic mean-field theory (single-mode condensate)}

We focus on zero temperature bosons and adopt the single-mode ansatz: all $N$ bosons occupy a single momentum $q=k_0+\phi$,
\begin{equation}
\langle \hat n_k\rangle = N\,\delta_{k,k_0},\qquad J=2N\sin q.
\nonumber 
\end{equation}
Using Eq.~\eqref{eq:Heff}, the energy per particle reads
\begin{equation}
{\,\frac{E(q)}{N} = -2C\cos q \;-\; \frac{2g^2}{\Delta}\,N\,\sin^2 q\, .\,}
\label{eq:En}
\end{equation}
The first term is the single-particle band energy (minimum at $q=0$ if $C>0$, at $q=\pi$ if $C<0$); the second term is the induced attractive ($\Delta>0$) or repulsive ($\Delta<0$) nonlinearity in $k$-space.

Minimizing Eq.~\eqref{eq:En} gives
\begin{equation}
\frac{d}{dq}\frac{E}{N} = \sin q\Big[2C - \frac{4g^2}{\Delta}N\cos q\Big]=0.
\nonumber 
\end{equation}
For $C>0$ the curvature at $q=0$ is
\begin{equation}
\left.\frac{d^2}{dq^2}\frac{E}{N}\right|_{q=0}
= 2C - \frac{4g^2}{\Delta}N,
\nonumber 
\end{equation}
which changes sign at
\begin{equation}
{\,N_c=\frac{C\,\Delta}{2g^2}\, .\,}
\nonumber 
%\label{eq:Nc}
\end{equation}
Hence,
\begin{align}
&N<N_c:\quad q=0 \ \text{(zero-current condensate)}, \nonumber \\
&N\ge N_c:\quad \cos q^*=\frac{N_c}{N},\ \ q^*\in(0,\pi/2], \nonumber 
\end{align}
with two symmetry-related minima at $\pm q^*$ breaking global QMF-current symmetry.

This is a finite-size effect. The scaling in the thermodynamic limit is $g \propto 1/\sqrt{N}$, $g_c=\sqrt{C \Delta/2}$, or 
\[
{\;(\delta n\,\theta)_c
= 2\,\arccos\!\left(\frac{-\Delta \pm \sqrt{\Delta^{2}+16}}{4}\right)\;}
\]
% Choose the sign/branch consistent with your physical constraints (e.g., g=\sin x \ge 0).

\subsection{Order parameter and critical behavior}

A natural order parameter is the current per particle,
\begin{equation}
\mu \equiv \frac{J}{2N}=\sin q.
\nonumber 
\end{equation}
In the ordered phase ($N\ge N_c$),
\begin{equation}
\mu_*=\sin q^*=\sqrt{1-\Big(\frac{N_c}{N}\Big)^2}\, .
\nonumber 
\end{equation}
Thus the transition is continuous (Ising-like in momentum space) with mean-field exponent $=1/2$.

\subsection{Ground-state energy}
For $N\le N_c$, $E_{\min}/N=-2C$. For $N\ge N_c$, inserting $\cos q^*=N_c/N$ into Eq.~\eqref{eq:En} yields
\begin{equation}
{\,\frac{E_{\min}}{N}
= -\,\frac{2g^2}{\Delta}\,N \;-\; \frac{C^2\,\Delta}{2g^2}\,\frac{1}{N}\, .\,}
\end{equation}

Expanding Eq.~\eqref{eq:En} around $q=0$ ($C>0$) gives
\begin{equation}
\frac{E}{N} = -2C
+\Big[C-\frac{2g^2}{\Delta}N\Big]q^2
+\Big[-\frac{C}{12}+\frac{2g^2}{3\Delta}N\Big]q^4+\cdots,
\end{equation}
so the quadratic coefficient changes sign at $N=N_c$ and the quartic coefficient at criticality is $C/4>0$, confirming a second-order pitchfork transition.

%%%%%%%%%%%%%%%%%%%%%%%%

\subsection{Finite-size quantization on a ring}

On a ring with $L$ sites, allowed values of momentum are $q_m=2\pi m/L$. The continuous transition is replaced by level crossings when a nonzero $q_m$ first becomes energetically favorable:
\begin{equation}
\Delta E_m/N = -2C(\cos q_m-1) - \frac{2g^2}{\Delta}N\sin^2 q_m<0,
\nonumber 
\end{equation}
which gives the stepped thresholds
\begin{equation}
N_c^{(m)}=\frac{C\Delta}{2g^2}\,\frac{1-\cos q_m}{\sin^2 q_m}
=\frac{C\Delta}{4g^2}\,\frac{q_m^2}{\sin^2(q_m/2)}
\;\xrightarrow{L\to\infty}\; N_c.
\nonumber 
\end{equation}

The validity of adiabatic elimination requires $|g J|\ll |\Delta|$. In the ordered phase $|J|\le 2N$, hence a conservative bound is 
\begin{equation}
{\,N \ll N_{\mathrm{ad}}\equiv \frac{\Delta}{2g}\, .\,}
\nonumber 
\end{equation}
A broad scaling window exists when $N_c\ll N_{\mathrm{ad}}$, i.e.\ $C\ll 2g$, which is fulfilled away from $\delta n\,\theta/2\approx 0,\pi$ where $g\to 0$.

\subsection{Experimental proxy via the qubit state}

The order parameter can be measured by performing the measurement of the qubit state. Indeed, before adiabatic elimination, the average  $\langle \sigma_z\rangle\simeq -(g/\Delta)\langle J\rangle$. In the ordered phase,
\begin{equation}
\langle \sigma_z\rangle \approx -\frac{2gN}{\Delta}\,\sin q^*
= -\frac{2gN}{\Delta}\,\sqrt{1-(N_c/N)^2},
\nonumber 
\end{equation}
which turns on continuously at $N_c$ and can serve as a readout of the order parameter.

\section{2D lattice model}

As an example of the 2D system, we study the Hofstadter model described by the Hamiltonian \cite{Hofstadter}
\begin{equation}
H = -t \sum_{\langle i, j \rangle} \left( e^{i \alpha_{ij}} c_i^\dagger c_j + \text{H.c.} \right),
\label{hofst1}
\end{equation}
where  the Peierls phase factor \(\alpha_{ij}\) is given by
$$
\alpha_{ij} = \frac{2\pi}{\phi_0} \int_{\mathbf{r}_i}^{\mathbf{r}_j} \mathbf{A} \cdot d\mathbf{r}, 
$$ and
\(\mathbf{A}\) is the vector potential corresponding to the magnetic field \(\mathbf{B}\).

%We consider the simplest case as before where we upgrade the gauge field to an operator,
As before, we consider $\hat{A}=IA_0+\hat{n} A$, where $n$ is the occupation number of the field, and we include only the two lowest levels so the two ladder phases can be set by $A_0$.  With a Hamiltonian for the two-level system described by $\sigma_{x}$, the total Hamiltonian will be 
\begin{equation} 
H = -t \sum_{\langle i, j \rangle} \left( e^{i \hat{n}\alpha_{ij}} c_i^\dagger c_j + \text{H.c.} \right)+\Delta \sigma_{x}\, . 
\label{hofst2}
\end{equation}

Note that the Peierls phases became operator-valued. One gets two block matrices with different fluxes coupled by $\Delta$ as in Eq.~\eqref{eq:f21}.
Using a pseudospin Hofstadter ansatz, we arrive at two coupled difference equations, each similar to the Harper equation in \cite{Hofstadter}. The two-leg Harper ladder is described by
\begin{align}
\psi^{(1)}_{n+1}+\psi^{(1)}_{n-1}
+2\cos(2\pi\alpha_1 n+k)\,\psi^{(1)}_n
+\Delta\,\psi^{(2)}_n
&=E\,\psi^{(1)}_n, \notag\\
\psi^{(2)}_{n+1}+\psi^{(2)}_{n-1}
+2\cos(2\pi\alpha_2 n+k)\,\psi^{(2)}_n
+\Delta\,\psi^{(1)}_n
&=E\,\psi^{(2)}_n.
\label{eq:coupled-harper}
\nonumber
\end{align}

Starting with matched fluxes $\alpha_1=\alpha_2\equiv\alpha$,
bonding/antibonding modes
$\phi^{(\pm)}_n=(\psi^{(1)}_n\pm\psi^{(2)}_n)/\sqrt2$
obey independent Harper equations,
\begin{equation}
\phi^{(\pm)}_{n+1}+\phi^{(\pm)}_{n-1}
+2\cos(2\pi\alpha n+k)\,\phi^{(\pm)}_n
=(E\mp\Delta)\,\phi^{(\pm)}_n,
\nonumber
\end{equation}
i.e., two copies of the Hofstadter spectrum shifted by $\pm\Delta$.
Now suppose that there is a small mismatch $\delta\alpha$ which might be a relevant case for experiments where the field producing this quantum phase is weak. For a small flux mismatch $|\delta\alpha|\ll 1$, the onsite potential difference
\[
\delta V_n \;\equiv\; V_2(n;k)-V_1(n;k) \;\neq\;0
\]
re-couples the bonding/antibonding channels in the $\pm$ basis. 

%%%%%%%%%%%%%%%%%%%%%%%%%

\subsection{Transfer matrix}

Defining
$\Phi_n=(\psi^{(1)}_n,\psi^{(1)}_{n-1},\psi^{(2)}_n,\psi^{(2)}_{n-1})^T$,
one iteration step $\Phi_{n+1}=T_n\Phi_n$ is determined by the matrix
\begin{equation}
T_n=
\begin{pmatrix}
E-V_1(n) & -1 & -\Delta & 0\\
1 & 0 & 0 & 0\\
-\Delta & 0 & E-V_2(n) & -1\\
0 & 0 & 1 & 0
\end{pmatrix},
\end{equation}
with $V_j(n)=2\cos(2\pi\alpha_j n+k)$. 
The smallest positive Lyapunov exponent
$\gamma(E)=\lim_{N\to\infty}\tfrac1N\log\sigma_i(T_{N-1}\cdots T_0)$
controls localization of the wave functions, where $\sigma_i$ is the singular value.
  
By symmetry the linear correction to the Lyapunov exponent $\gamma(E)$ vanishes, and the leading contribution is quadratic in the small parameters. We find
\begin{equation}
\gamma(E)
= c(E)\,|\Delta|^2\,|M(\delta\alpha)|^2
+ \mathcal O\!\Big(|\Delta|^2|\delta\alpha|^3,\,|\Delta|^4\Big),
\label{eq:gamma-small-mismatch}
\end{equation}
where the {\it dephasing structure factor} 
\begin{equation}
M(\delta\alpha) \;\equiv\;
\lim_{N\to\infty}\frac{1}{N}\sum_{n=0}^{N-1}
e^{i2\pi\delta\alpha n}\,w_n
\label{eq:M-dalpha}
\end{equation}
weighs the normalized site amplitudes $w_n$ of the $\pm$ channels.  
The prefactor $c(E)$ is model-dependent and encodes the local spectral curvature (e.g., group velocities or Green’s function weights) of the decoupled Harper chains. The quasi-periodic and inter-leg dephasing turns on a finite Lyapunov exponent. 

Equation~\eqref{eq:gamma-small-mismatch} shows that $\gamma(E)$ grows quadratically with both $|\Delta|$ and $|\delta\alpha|$, reflecting the fact that perfect phase alignment at $\delta\alpha=0$ forbids exponential growth (criticality). Any persistent $n$-dependent phase mismatch induces weak backscattering between the $\pm$ channels, thereby turning on a small but finite Lyapunov exponent. As $|\delta\alpha|$ increases, $\gamma(E)$ rises rapidly, with pronounced \emph{revivals} near small-denominator commensurations where $|M(\delta\alpha)|$ is enhanced. So there is a crossover from metallic to insulator phases at nonzero $|\delta\alpha|$. 
 
%%%%%%%%%%%%%%%%%%%%%%%%%%%%

\subsection{Periodicity and band counting}

For rational $\alpha_j=p_j/q_j$ the coupled ladder has a larger "supercell" with a period equal to the least common multiple of the individual periods, 
$L=\mathrm{lcm}(q_1,q_2)$, yielding $2L$ magnetic subbands at each Bloch momentum. The system is insulating in the sense that the eigenstates are localized, if either of the two $\alpha$ is irrational, and conducting when they are commensurable.

\begin{figure}[t]               % placement “t” for top; use h or b if preferred
  \centering
  %--- sub-figure (a) ---
  \subfloat[]{%
    \includegraphics[width=0.50\linewidth]{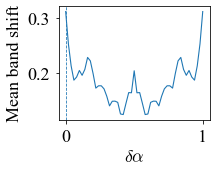}%
    \label{fig:panel_a}}%
  \hfill
  %--- sub-figure (b) ---
  \subfloat[]{%
    \includegraphics[width=0.48\linewidth]{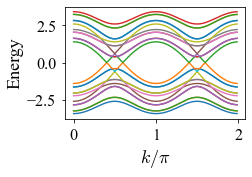}%
    \label{fig:panel_b}}%
  \hfill
  %--- sub-figure (c) ---
   \subfloat[]{%
    \includegraphics[width=0.480\linewidth]{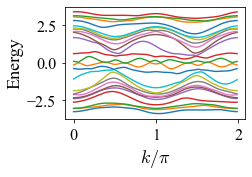}%
    \label{fig:panel_b}}%
     \hfill
%---------------D----------
   \subfloat[]{%
    \includegraphics[width=0.480\linewidth]{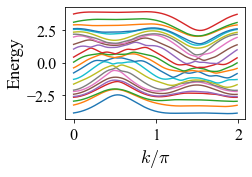}%
    \label{fig:panel_b}}%
     \hfill

  \caption{(a) Hybridization strength as a function of flux mismatch 
$\delta\alpha=\alpha_2-\alpha_1$. 
This metric is defined as the mean spectral shift between the coupled 
($\Delta\neq0$) and uncoupled ($\Delta=0$) spectra, averaged over momentum $k$. 
The plot parameters are $\alpha_1= 0.5$, $\Delta=0.6$, and the magnetic unit cell $n$ varies from 1 to 12 in the Harper equations. 
(b) Energy bands normalized by $t$ as a function of wavenumber $k$ for $\alpha_1=\alpha_2=0.5$ (c) Same for $\alpha_1=0.5$ and $\alpha_2=0.7$ (d) Same for $\alpha_1=0.5$ and $\alpha_2=0.95$. }
    
  \label{hybrid}
\end{figure}

Figure \ref{hybrid}(a) illustrates the hybridization strength as a function of flux mismatch 
$\delta\alpha=\alpha_2-\alpha_1$, where the former 
 is defined as the mean spectral shift between the coupled 
($\Delta\neq0$) and uncoupled ($\Delta=0$) energy spectra, averaged over momentum $k$. 
A sharp maximum occurs at $\delta\alpha=0$, where the two Hofstadter blocks are identical 
and rung coupling $\Delta$ produces maximal hybridization. 
The hybridization decays rapidly with $|\delta\alpha|$, reflecting dephasing 
between the two magnetic Bloch waves, with mild revivals near 
commensurate flux differences. The plots in Fig.~\ref{hybrid}(b)-(d) show the bandstructure in the case of (b) strongest hybridization  when  $\delta \alpha$ is zero so the mean shift is maximal, (c) intermediate value of the mean shift when $\delta \alpha = 0.2$, and (d) the lowest mean shift when $\delta \alpha = 0.45$.  

%%%%%%%%%%%%%%%%%%%%%%%%%%

\subsection{Perturbation theory for small $\delta\alpha$}

For $\alpha_2=\alpha_1+\delta\alpha$ one finds
$V_2(n)=V_1(n)+\delta\alpha\,V'_1(n)+\dots$.
If we write $H=H_0+W$, then $H_0$ is block-diagonalized by the $\pm$ basis
with eigenvalues $\varepsilon_m(\alpha_1)\pm\Delta$.
First-order corrections vanish by symmetry, while the leading
second-order shift reads
\begin{equation}
E_{m,\pm}\simeq \varepsilon_m(\alpha_1)\pm\Delta
+\frac{\delta\alpha^2}{4}\sum_{m'}
\frac{|\langle m|V'_1|m'\rangle|^2}
{(\varepsilon_m-\varepsilon_{m'})^2+(2\Delta)^2}.
\nonumber 
\end{equation}
Thus hybridization scales as $|\Delta|^2|\delta\alpha|^2$ at small mismatch,
with avoided-crossing gaps
$\Delta E\simeq 2|\Delta\langle u^{(1)}|u^{(2)}\rangle|$
controlled by wavefunction overlap.

We solve the coupled Harper equations numerically and plot the eigenenergies in Fig.~\ref{fig:hofst2},  fixing $\alpha_1=1/3$ and sweeping over rational $\alpha_2$ for $\Delta=t/2$. Notice the sensitivity to the value of the flux difference $\delta\alpha$ between the two QMF states.

\begin{figure}[htbp]
  \centering
  % First row: two figures
  %\begin{subfigure}{\textwidth}
    \includegraphics[width=\linewidth]{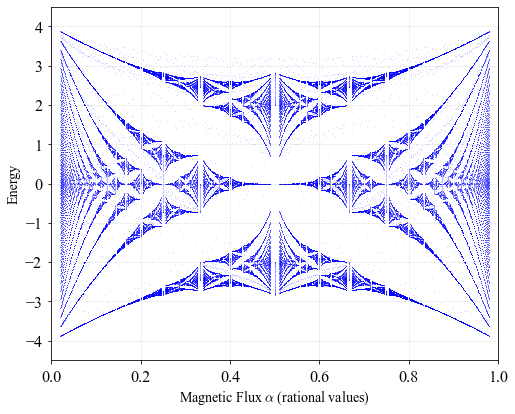}
    \caption{Eigenenergies normalized by $t$ in the original Hofstadter butterfly described by the Hamiltonian Eq.~(\ref{hofst1}). }
      \label{fig:hofst1}
  \end{figure}

\begin{figure}[htbp]
    \includegraphics[width=\linewidth]{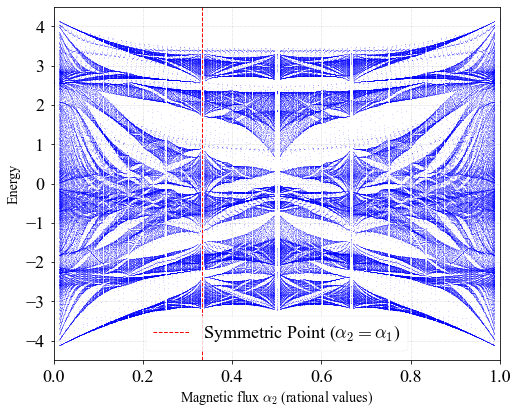}
    \caption{Eigenenergies normalized by $t$ in the model described by the Hamiltonian Eq.~(\ref{hofst2}), with two states of QMF and tunneling $\Delta$ between them: $\Delta=0.6 t$, $ \alpha_1=1/3$, and we sweep over rational $\alpha_2$. If both $\alpha$'s are rational numbers then there is a common period resulting in a fractal structure. If one of them is irrational, then the eigenfunctions are localized; any infinitesimal detuning from commensuration makes states localized. For a very irrational $\alpha_1$ equal to the golden ratio   
    the spectrum is the result of solving the two coupled Harper equations. 
    }
  %\end{subfigure}
  
  \label{fig:hofst2}
\end{figure}

The original Hofstadter butterfly is plotted in Fig.~\ref{fig:hofst1} for comparison. 
Topological phase transitions can occur under variations of $\Delta$ and $\delta\alpha$ as gaps will open and close.

  %\hfill

Note that one could imagine a  one-particle picture that does not require any QMF: for example,  a two-layer system with equal tunneling amplitude between any site of one layer and every site of the other one will realize this Hamiltonian.

Finally, upon generalization to multiple flux qubits on a lattice, the system bears close resemblance to quantum link models, where lattice gauge fields are represented by finite-dimensional Hilbert spaces (e.g., spin-1/2 operators on links) \cite{Tagliacozzo2013}. Our use of a two-level flux qubit as a quantized vector potential directly maps onto this framework. The model can simulate $Z_2$ lattice gauge theories and potentially realize topologically ordered phases such as the toric code.

\section{Implementation of artificial QMF}

\subsection{Superconducting qubits}

The Hamiltonians above can be simulated using SC qubits as a synthetic QMF background; not to be confused with the case of using them as a source of real QMF. 
Consider a planar square lattice of fixed-frequency SC qubits (for example transmons) $\{j\}$ coupled by flux-tunable SQUID couplers that are parametrically driven to engineer complex exchange amplitudes.
In the rotating frame (and within the single-excitation sector), qubit flips $\sigma_j^\pm$ realize a tight-binding model.
We upgrade the Peierls phase on each link to be operator valued by weakly flux-biasing the coupler with a dedicated two-level gauge qubit (e.g., a flux qubit) per plaquette $p$, so that the link acquires a phase shift set by the gauge state:
\begin{align}
%\label{eq:H_SC_2D_gauge}
H=\sum_{\langle i j\rangle}\!\Big[J_{ij}\,e^{\,i\!\left(A^{(0)}_{ij}+g_{ij}\,\hat{\sigma}^{(p_{ij})}_z\right)}\,\sigma_i^+ \sigma_j^-+\text{h.c.}\Big] \nonumber \\
+\sum_p\Big(\tfrac{\Delta_p}{2}\hat{\sigma}^{(p)}_x+\tfrac{\varepsilon_p}{2}\hat{\sigma}^{(p)}_z\Big).
\nonumber 
\end{align}
Here $A^{(0)}_{ij}$ program a background synthetic magnetic field through parametric-driven phases on the couplers, while the values of $g_{ij}$ are set by a calibrated mutual inductance $M$ between the coupler loop and the gauge qubit’s persistent current $I_p$ (or an equivalent dispersive-phase lever arm).
Choosing identical $g_{ij}\equiv g$ on the four links of the plaquette $p$ makes the net flux \emph{qubit conditioned}, $\Phi_p=\Phi^{(0)}\pm 4g$, with the sign determined by the eigenvalue of $\hat{\sigma}^{(p)}_z$.
Preparing $\lvert g\rangle_p$ versus $\lvert e\rangle_p$ flips the chirality of edge modes and the sign of equilibrium currents; superpositions of the gauge qubit generate matter–gauge entanglement visible in Ramsey interferometry on the gauge and in link-current correlators measured dispersively on the array.
This architecture leverages (i) high-coherence circuit-QED devices and low-disorder 2D resonator/qubit lattices; (ii) tunable/parametric couplers whose amplitude and phase can be set with nanosecond resolution; and (iii) synthetic magnetic fields for microwave excitations in SC lattices, demonstrated, e.g., in \cite{BlaisRMP2021,Underwood2012,Chen2014,Caldwell2018,Mundada2019,SungPRX2021,Roushan2017}.

To implement only two states of the global QMF the system can be designed such that every link coupling is generated by the same parametric pump.
Routing a single microwave tone to all flux-tunable couplers through a balanced splitter tree makes the Peierls phase \(\phi\) common to every link.
We render the gauge \emph{quantum} by inserting a qubit-conditioned phase shifter ahead of the splitter: reflecting (or transmitting) the pump through a resonator dispersively coupled to a dedicated gauge qubit imprints a state-dependent phase \(\phi=\phi_{0}+\phi_{1}\hat{\sigma}_{z}\), a standard circuit-QED mechanism \cite{BlaisRMP2021,ClerkRMP2010}.
In the rotating frame, the array implements the Hamiltonian 
\begin{equation}
\label{eq:H_global_gauge}
H=\sum_{\langle ij\rangle} J\,e^{\,i\!\left(A_{0}+ \phi_{1}\hat{\sigma}_{z}\right)}\,\sigma_i^{+}\sigma_j^{-}+\text{h.c.}
+\frac{\Delta}{2}\hat{\sigma}_{x}+\frac{\varepsilon}{2}\hat{\sigma}_{z},
\nonumber 
\end{equation}
so that preparing the gauge in \(|g\rangle\) or \(|e\rangle\) selects one of two \emph{global} flux configurations, while a superposition generates matter–gauge entanglement.
Uniform phase delivery exploits established parametric/tunable couplers whose complex exchange amplitudes are set by pump amplitude and phase \cite{Chen2014,Caldwell2018,Mundada2019,SungPRX2021}.
With a background programmed flux pattern \(A_{0}\), this scheme yields two globally distinct gauge sectors for chiral transport and equilibrium currents, extending earlier demonstrations of synthetic magnetic fields in SC lattices \cite{Underwood2012}.

A more involved way to implement the two global states is to use a global flux-bias bus.
As a DC route, a superconducting \emph{flux bus} can be weakly magnetically coupled to every coupler SQUID.
A single flux qubit with persistent current \(\pm I_{p}\) then induces a uniform offset \(\delta\Phi\) on all couplers, shifting their coupling phase \(A_{ij}\!\to\!A_{ij}+g\,\hat{\sigma}_{z}\).
Practical implementation favors gradiometric couplers and per-link m\(\Phi_{0}\)-scale trims to equalize residual mutual-inductance variations; however, achieving sizable, uniform \(\delta\Phi\) across large arrays is generally more challenging than the pump-phase approach, which directly leverages microwave-phase uniformity and cQED dispersive control \cite{BlaisRMP2021,KrantzAPR2019}.

Examples of the SC systems with an artificial gauge field include \cite{Roushan2017,oliver}. The synthetic gauge field is implemented in these setups via Flouqet engineering by modulating the interaction qubits. Using quantum states of light instead of classical light to modulate the coupling of the qubits will realize the QMF.

\subsection{Waveguide and cavity QED}

In the three-cavity scheme of \cite{WangScullyPRL2016}, periodic modulation of the cavity frequencies with relative phases produces a synthetic magnetic flux on a triangular photonic loop; remarkably, the chirality of photon circulation depends on the internal state of the single atom.
Promoting the atom to an explicit dynamical degree of freedom, driving the atoms or changing the atom Hamiltonian turns the Peierls phase into an operator-valued gauge conditioned on the atom state:
\begin{multline}
\label{eq:H_triangle_quantum_gauge}
H=\sum_{j=1}^{3}\omega_c a_j^\dagger a_j+\frac{\omega_a}{2}\hat\sigma_z
+\sum_{j} g_j\!\left(a_j^\dagger \hat\sigma_-+a_j \hat\sigma_+\right)\\
+\sum_{j} J\,e^{\,i\!\left(\Phi_0+\Phi_1\hat\sigma_z\right)} a_{j+1}^\dagger a_j+\text{h.c.},
\end{multline}
with $a_{4}\!\equiv\!a_1$ and where $\Phi_0$ is the programmed (classical) flux from modulation, while $\Phi_1\hat\sigma_z$ encodes the atom-conditioned contribution. 
Two immediate consequences follow:  
(i) The realization of two gauge sectors, where preparing the atom in a $\lvert g\rangle$ or $\lvert e\rangle$ state elects opposite effective fluxes $\Phi_{\pm}=\Phi_0\pm\Phi_1$, yielding chiral ground-state currents $\langle \hat{J}_\circlearrowleft\rangle\!\propto\!\sin\Phi_{\pm}$ with  opposite circulation, as in the QED synthetic magnetic field cavity \cite{WangScullyPRL2016,MalekiPRA2023}. 
(ii) Matter–gauge entanglement in which an atomic superposition $(\lvert g\rangle+\lvert e\rangle)/\sqrt2$ entangles with circulating photonic eigenmodes; interference between the two flux sectors suppresses mean chirality while enhancing current fluctuations, and measurement of the atom projects the photons into opposite-direction circulating states to the mesoscopic cat-state dynamics reported in the three-cavity setup.  

Beyond the dispersive limit, hybridization yields \emph{polaritonic} bands on the triangle, so the synthetic flux acts on light–matter excitations; with increasing drive or excitation number, the Jaynes–Cummings nonlinearity produces photon-number–dependent phase shifts and can trigger chiral superradiant phases in ring generalizations \cite{LiPRA2023}. 
Related ring-cavity analyses further show that cavity QED can realize strong synthetic gauge fields and spin–orbit couplings for atoms, supporting the operator-valued generalization above \cite{MivehvarPRA2014}.

\subsection{Synthetic dimensions}

Artificial gauge fields in synthetic dimensions exploit internal degrees of freedom (spin, hyperfine states, mode frequencies, etc.) as a dimension for lattice sites,  whereas phase-engineered couplings induce a magnetic vector potential. Simulation of higher-dimensional and topological phenomena, such as quantum Hall physics and even 4D quantum Hall effect has been realized in experimental platforms including ultracold atoms, photonic resonators, and SC circuits \cite{Boada2012,Celi2014,Yuan2018,Goldman2016}.

\subsubsection{Superconducting circuits}
For SC circuits, a 2D lattice in which the real axis is a chain of microwave modes (resonators or qubits) indexed by $j$, while the synthetic axis is provided by a ladder of internal levels or distinct modal frequencies $\{\lvert s\rangle\}$ coupled by parametrically driven junctions can be realized. 
Nearest–neighbour synthetic hops are implemented by modulating a SQUID/coupler at the intermode detuning; the drive phase sets the Peierls phase, yielding complex tunneling $\kappa_s e^{i\varphi}$.
The Hamiltonian reads
\begin{multline}
\label{eq:H_cQED_synth_qubit}
H=\sum_{j,s}\Big[
t_x\,a_{j+1,s}^\dagger a_{j,s}
+\kappa_s\,e^{\,i\!\left(\varphi_0+\varphi_1\hat{\sigma}_z\right)}\,
a_{j,s+1}^\dagger a_{j,s}+\text{h.c.}\Big] \\
+\;\frac{\Delta}{2}\hat{\sigma}_x
+\frac{\varepsilon}{2}\hat{\sigma}_z \, .
\end{multline}
Here, $\hat{\sigma}_{x,z}$ act on an ancillary gauge qubit that imprints an operator-valued Peierls phase on synthetic hops (via dispersive control of the pump phase or a flux-biased coupler), so that preparing $\lvert g\rangle$ vs. $\lvert e\rangle$ selects opposite flux offsets, while superpositions generate matter–gauge entanglement.
This approach leverages established circuit-QED control and parametric couplings for realizing synthetic magnetic fields and topological band structures in superconducting lattices \cite{BlaisRMP2021,Underwood2012,Roushan2017,WangNPJQI2016}.

For photonic frequency-lattices implementing a qubit-conditioned synthetic gauge field might be acheivable. A SC qubit in circuit QED can readily imprint a state-dependent phase on a weak microwave probe via dispersive coupling; that phase serves as the gauge “knob” \cite{BlaisRMP2021,ClerkRMP2010}. The phase-tagged probe is then raised to the necessary power without destroying phase information using quantum-limited Josephson amplifiers (JPA/JTWPA) \cite{CastellanosBeltran2008,Macklin2015}. Driving the ring-resonator electro-optic modulator (EOM) with this tone realizes a Peierls phase on frequency-mode hopping; hence the EOM drive phase becomes the synthetic gauge potential \cite{YuanOL2016,Dutt2019}. The feasibility is supported by cryogenic superconducting–electro-optic integration in thin-film lithium niobate, which demonstrates co-integration with SC  microwave resonators and bidirectional microwave–optical conversion, establishing a practical route for qubit-controlled EOM phases at millikelvin temperatures \cite{McKenna2020,Xu2021}.

\subsubsection{Ultracold atoms with Raman synthetic ladders}
A synthetic axis formed by hyperfine states $\{\lvert m_F\rangle\}$ of the atoms can be realized.
Neighboring synthetic sites are coupled by two-photon Raman transitions with complex tunneling $t_s e^{i\phi(x)}$, where the Raman phase $\phi(x)=\Delta\mathbf{k}\!\cdot\!\mathbf{r}$ yields a uniform magnetic flux per plaquette $\Phi$ of the $x\times$synthetic ladder \cite{Celi2014,CooperRMP2019,Dalibard2011}.
Hard edges along the synthetic direction generate chiral edge channels, and chiral motion has been observed directly in Raman-engineered Hall ribbons for both fermions and bosons \cite{Mancini2015,Stuhl2015}.
Within this framework, one can implement an operator-valued Peierls phase by conditioning the Raman phase on a two-state gauge degree of freedom:
\begin{align}
\label{eq:H_raman_synth_qubit}
H=\sum_{j,s}\Big[t_x\,c_{j+1,s}^\dagger c_{j,s}
+\frac{\Omega}{2}\,e^{\,i(\phi_0+\phi_1\hat{\sigma}_z)}\,c_{j,s+1}^\dagger c_{j,s}
+\text{h.c.}\Big] \nonumber \\
+\frac{\Delta}{2}\hat{\sigma}_x+\frac{\varepsilon}{2}\hat{\sigma}_z \, .
\nonumber 
\end{align} 
Preparing the gauge qubit in a $\lvert g\rangle$ or $\lvert e\rangle$ state selects opposite flux offsets, while superposition states generate matter--gauge entanglement that can be detected via state-resolved time-of-flight and qubit Ramsey interferometry \cite{Celi2014,Livi2016,CooperRMP2019}.
Background on synthetic-dimension ladders and Raman control is reviewed in \cite{Celi2014,Mancini2015,Stuhl2015,Livi2016,CooperRMP2019,Dalibard2011}.

\subsubsection{Trapped ions}
In a 1D chain of trapped ions a 2D synthetic lattice can be realized using one real spatial axis (ion index $j$) and a synthetic axis spanned either by internal spin states or by motional Fock states $\{\lvert s\rangle\}$.
State-dependent stimulated-Raman (or microwave-gradient) drives generate nearest–neighbour synthetic hops with controlled complex amplitude $\Omega_s e^{i\phi}$, where the optical phase difference sets the Peierls phase.
To promote the gauge to a dynamical two-state degree of freedom, one conditions one Raman leg on a  \emph{gauge ion} (or auxiliary qubit) to realize
\begin{align}
%\label{eq:H_ion_synth_qubit}
H=\sum_{j,s}\Big[J\,b_{j+1,s}^\dagger b_{j,s}
+\frac{\Omega_s}{2}\,e^{\,i\!\left(\phi_0+\phi_1\hat{\sigma}_z\right)}\,b_{j,s+1}^\dagger b_{j,s}
+\text{h.c.}\Big] \nonumber \\
+\frac{\Delta}{2}\hat{\sigma}_x+\frac{\varepsilon}{2}\hat{\sigma}_z \, ,
\nonumber 
\end{align}
so that the gauge-ion state controls the synthetic flux.
This construction builds on trapped-ion schemes where spin–motion couplings engineer synthetic dimensions and artificial magnetic flux, and on demonstrations of Harper–Hofstadter ladders with measurable chiral currents \cite{Bermudez2011,WangPRL2024,ArguelloLuengo2024}.

As another example, in \cite{Ion2024} a laser illuminating a single ion simulated the Hofstadter system. Using a quantum state of light instead of a laser can encode the QMF. For example,  a Fock state $|0\rangle+|N\rangle$ provides two quantum DOFs for the synthetic magnetic flux. Illuminating with a squeezed state of light is another way to introduce QMF in this system.  

\subsection{Quantum lattice geometry}

For electronic systems, the lattice geometry determines the effective potential experienced by the electrons, and thereby controls both the band structure and the topological properties of the bands. Lattice geometry can give rise to an artificial magnetic field: a notable example is strained graphene where nonuniform lattice deformations generate artificial magnetic fields that can reach extremely large effective field strengths.

The most straightforward realization would be a bilayer system composed of two 2D materials, such as graphene, where each layer experiences a different strain profile, and the layers are coupled by an interlayer tunneling amplitude $\Delta$. For a single-particle scenario, this configuration can simulate our target Hamiltonian. However, because the strain in each layer is not dynamical quantum variables, the system cannot reproduce the many-body Hamiltonian in which particles interact via energy exchange with an artificial QMF.

To realize an artificial QMF on the lattice it has to be in an unusual macroscopic superposition of lattice configurations. A possible route is to make the 2D material part of a resonator or bond it to a high-Q membrane. In this scheme, the quantum state of the strain field would be coupled to the quantum state of the resonator, which could be prepared in a coherent superposition, thereby enabling fully dynamical QMF mediated interactions between the electrons in the material 
\cite{Barzanjeh2022Optomechanics}.

\section{Conclusions}

We have shown that when the real or synthetic magnetic flux is promoted to a quantum dynamical DOF, novel nonlocal interaction effects and topological phenomena can arise in otherwise non-interacting systems. In the 1D rings encircling a common QMF, particles on a ring are coupled to the QMF, which gives rise to an emergent nonlocal interaction between the particles, with tunable strength and interesting nonlinear effects that can result in a phase transition. For a 2D particle system in a QMF, there is a much richer playground of nonlocal interactions with controllable topological properties, tunable nonlinearities, and crossover between the ground states of different chirality. In addition to that the interaction has fundamental physics interest as it lakes spacial dependence.

Systems with synthetic gauge fields provide the platforms to realize the same physics with the advantage of tunability enabling much higher strength of the synthetic QMF and the interaction of neutral particles.  In particular, there is possibility to realize strongly interacting photonic systems without direct light-matter coupling. Our work provides a theoretical pathway to realize tunable nonlinear interactions purely from the quantization of synthetic gauge fields, without requiring intrinsic optical nonlinearities. These case studies illustrate a general principle: quantum fluctuations of QMF can induce effective interactions that have no counterpart in classical fields.

Our findings can be implemented across different platforms in the light of recent experiments and proposals which offer a rich toolbox of systems with synthetic dimensions and implementations of synthetic gauge fields. 

For future directions, by coupling multiple lattice plaquettes through a single or multiple QMF (as in a circuit QED architecture), one might generate long-range correlated hopping and interaction patterns that stabilize exotic phases such as anyonic liquids or dynamical gauge field theories in the quantum regime. Ultimately, merging quantum optics with topological matter might enable the creation of quasi-particles that are hybrids of light, matter, and gauge fields, thus opening avenues to observe phenomena like photon-induced topological phase transitions and perhaps even analogs of high-energy physics processes in tabletop experiments.

\section{Acknowledgments} 

This work has been supported in part by the Laboratory Directed Research and Development program and Sandia University Partnerships Network ( SUPN) program and performed in part at the Center for Integrated Nanotechnologies ( CINT), an Office of Science User Facility operated for the U.S. Department of Energy (DOE) Office of Science. Sandia National Laboratories is a multimission laboratory managed and operated by National Technology and Engineering Solutions of Sandia, LLC., a wholly owned subsidiary of Honeywell International, Inc., for the U.S. Department of Energy's National Nuclear Security Administration under Contract No. DE-NA0003525. This article describes objective technical results and analysis. Any subjective views or opinions that might be expressed in the article do not necessarily represent the views of the U.S. Department of Energy or the United States Government.

%%%%%%%%%%%%%%%%%%%%%%%%%%%%%%%%%%%%%%%%%%%%%

\bibliographystyle{apsrev4-2}
\bibliography{References}

\end{document}